\documentclass[dvipsnames,twocolumn]{aastex631}

\usepackage{placeins}
\usepackage{graphicx}
\usepackage{amsmath}
\usepackage{txfonts}
\usepackage{comment}
\usepackage{color}

\newcommand{\rgas}[0]{\ensuremath{R_{\rm CO,\ 90\%}}}
\newcommand{\rgasobs}[0]{\ensuremath{R^{\rm obs}_{\rm CO,\ 90\%}}}
\newcommand{\rgasmodel}[0]{\ensuremath{R^{\rm model}_{\rm CO,\ 90\%}}}
\newcommand{\rdust}[0]{\ensuremath{R_{\rm dust,\ 90\%}}}
\newcommand{\rdustfrank}[0]{\ensuremath{R^{\rm FRANK}_{\rm dust,\ 90\%}}}

\newcommand{\co}[0]{\ensuremath{^{12}\mathrm{CO}}}
\newcommand{\xco}[0]{\ensuremath{^{13}\mathrm{CO}}}
\newcommand{\cyo}[0]{\ensuremath{\mathrm{C}^{18}\mathrm{O}}}

\begin{document}
%\nolinenumbers

\title{The ALMA Survey of Gas Evolution of PROtoplanetary Disks (AGE-PRO): XI.  Beam-corrected gas disk sizes from fitting \co\ moment zero maps}

\correspondingauthor{Leon Trapman}
\email{ltrapman@wisc.edu}
\author[0000-0002-8623-9703]{Leon Trapman}
\affiliation{Department of Astronomy, University of Wisconsin-Madison, 
475 N Charter St, Madison, WI 53706, USA}

\author[0000-0002-4147-3846]{Miguel Vioque}
\affiliation{European Southern Observatory, Karl-Schwarzschild-Str. 2, 85748 Garching bei München, Germany}
\affiliation{Joint ALMA Observatory, Alonso de Córdova 3107, Vitacura, Santiago 763-0355, Chile}

\author[0000-0002-2358-4796]{Nicol\'as T. Kurtovic}
\affiliation{Max Planck Institute for Extraterrestrial Physics, Giessenbachstrasse 1, D-85748 Garching, Germany}
\affiliation{Max-Planck-Institut fur Astronomie (MPIA), Konigstuhl 17, 69117 Heidelberg, Germany}

\author[0000-0002-0661-7517]{Ke Zhang}
\affiliation{Department of Astronomy, University of Wisconsin-Madison, 
475 N Charter St, Madison, WI 53706, USA}

\author[0000-0003-4853-5736]{Giovanni P. Rosotti}
\affiliation{Dipartimento di Fisica, Università degli Studi di Milano, Via Celoria 16, I-20133 Milano, Italy}

\author[0000-0001-8764-1780]{Paola Pinilla}
\affiliation{Mullard Space Science Laboratory, University College London, 
Holmbury St Mary, Dorking, Surrey RH5 6NT, UK}

\author[0000-0003-2251-0602]{John Carpenter}
\affiliation{Joint ALMA Observatory, Alonso de Córdova 3107, Vitacura, Santiago 763-0355, Chile}

\author[0000-0002-2828-1153]{Lucas A. Cieza}
\affiliation{Instituto de Estudios Astrofísicos, Universidad Diego Portales, Av. Ejercito 441, Santiago, Chile}

\author[0000-0001-7962-1683]{Ilaria Pascucci}
\affiliation{Lunar and Planetary Laboratory, the University of Arizona, Tucson, AZ 85721, USA}

\author[0009-0004-8091-5055]{Rossella Anania}
\affiliation{Dipartimento di Fisica, Università degli Studi di Milano, Via Celoria 16, I-20133 Milano, Italy}

\author[0000-0002-7238-2306]{Carolina Agurto-Gangas}
\affiliation{Departamento de Astronom\'ia, Universidad de Chile, Camino El Observatorio 1515, Las Condes, Santiago, Chile}

\author[0000-0003-0777-7392]{Dingshan Deng}
\affiliation{Lunar and Planetary Laboratory, the University of Arizona, Tucson, AZ 85721, USA}

\author[0000-0002-1575-680X]{James Miley}
\affiliation{Departamento de Física, Universidad de Santiago de Chile, Av. Victor Jara 3659, Santiago, Chile}
\affiliation{Millennium Nucleus on Young Exoplanets and their Moons (YEMS), Chile} 
\affiliation{Center for Interdisciplinary Research in Astrophysics and Space Exploration (CIRAS), Universidad de Santiago, Chile}

\author[0000-0002-1199-9564]{Laura M. P\'erez}
\affiliation{Departamento de Astronom\'ia, Universidad de Chile, Camino El Observatorio 1515, Las Condes, Santiago, Chile}

\author[0000-0002-5991-8073]{Anibal Sierra}
\affiliation{Departamento de Astronom\'ia, Universidad de Chile, Camino El Observatorio 1515, Las Condes, Santiago, Chile}
\affiliation{Mullard Space Science Laboratory, University College London, 
Holmbury St Mary, Dorking, Surrey RH5 6NT, UK}

\author[0000-0002-1103-3225]{Beno\^it Tabone }
\affiliation{Universit\'e Paris-Saclay, CNRS, Institut d'Astrophysique Spatiale, Orsay, France}

\author[0000-0003-3573-8163]{Dary A. Ru\'iz-Rodr\'iguez}
\affiliation{National Radio Astronomy Observatory, 520 Edgemont Rd., Charlottesville, VA 22903, USA}

\author[0000-0003-4907-189X]{Camilo Gonz\'alez-Ruilova}
\affiliation{Instituto de Estudios Astrofísicos, Universidad Diego Portales, Av. Ejercito 441, Santiago, Chile}
\affiliation{Millennium Nucleus on Young Exoplanets and their Moons (YEMS), Chile}
\affiliation{Center for Interdisciplinary Research in Astrophysics and Space Exploration (CIRAS), Universidad de Santiago, Chile}

\author[0000-0001-9961-8203]{Estephani TorresVillanueva}
\affiliation{Department of Astronomy, University of Wisconsin-Madison, 
475 N Charter St, Madison, WI 53706, USA}

%%% shortened abstract [max 250 words]
\begin{abstract}
The inward drift of mm-cm sized pebbles in protoplanetary disks has become an important part of our current theories of planet formation and, more recently, planet composition as well. The gas-to-dust size ratio of protoplanetary disks can provide an important constraint on how pebbles have drifted inward provided that observational effects, especially resolution, can be accounted for.
Here we present a method for fitting beam-convolved models to integrated intensity maps of line emission using the \texttt{astropy} python package and use it to fit \co\ moment zero maps of ten Lupus and ten Upper Scorpius protoplanetary disks from the AGE-PRO ALMA Large Program, a sample of disks around M3-K6 stars that cover the $\sim1$ to 6 Myr of gas disk evolution. From the unconvolved best fit models we measure the gas disk size (\rgasmodel), which we combine with the dust disk size (\rdustfrank) from continuum visibility fits from \cite{AGEPRO_X_substructures} to compute beam-corrected gas-to-dust size ratios.
In our sample we find gas-to-dust size ratios between $\sim1$ and $\sim5.5$, with a median value of $2.78^{+0.37}_{-0.32}$. Contrary to models of dust evolution that predict an increasing size ratio with time, we find that the younger disks in Lupus have similar (or even larger) median ratios $(3.02^{+0.33}_{-0.33})$ than the older disks in Upper Sco $(2.46^{+0.53}_{-0.38})$. A possible explanation to this discrepancy is that pebble drift is halted in dust traps combined with truncation of the gas disk by external photo-evaporation in Upper Sco, although survivorship bias could also play a role.
\end{abstract}

%
%-------------------------------------------------------------------
\section{Introduction}
\label{sec: introduction}

Our knowledge of exoplanets has increased tremendously over the last two decades, with over 5000 exoplanets detected\footnote{\url{https://exoplanetarchive.ipac.caltech.edu/}} in a wide variety of system architectures (see, e.g. \citealt{Lissauer2023,Weiss2023}). 
Similarly, our understanding of how these planetary systems formed in disks of gas and dust around young stars has also expanded considerably (see, e.g., \citealt{Drazkowska2023}). 
Currently favored among planet formation theories is pebble accretion, where planetary embryos are able to rapidly accrete their mass by accreting millimeter-centimeter size pebbles that are expected to drift inwards from the outer part of the protoplanetary disks. 
In addition to building up planets, recent ALMA and JWST results suggest that these pebbles can also significantly alter the composition of the gas in the inner disk (e.g. \citealt{Zhang2020a,Banzatti2023,Gasman2023,Xie2023}). 

The difference between the gas disk size and the dust disk size of protoplanetary disks is an important indirect measure of pebble drift. The rationale behind this is that the gas disk size represents the starting point of the pebbles and the dust disk size represents their current maximum extent from the star. The gas-to-dust size ratio then gives us a measure of how far the pebbles have drifted inward. If measured at different points in the disk lifetime, or conversely for a set of disks with different ages, this would gives us a measure of the typical pebble drift in protoplanetary disks. 
\cite{Long2022} carried out this experiment with a sample of 44 disks and found no clear evolution with age, but their sample was heavily weighted towards disks $\sim2-3$ Myr old. In addition, there is a stellar dependence expected as pebbles drift more efficiently around very low mass stars \citep{Pinilla2013,Pinilla2022}.

Observationally measuring and interpreting the gas-to-dust size ratio brings with it several complications. The gas in disks is dominated by H$_2$, which does not emit significantly at the cold temperatures of the bulk disk material (e.g. \citealt{Thi2001,Carmona2011}). The gas disk size is therefore commonly measured from optically thick \co\ rotational emission which remains bright out to the low densities found in the outer disk. 
This is in contrast to optically thinner isotopologue lines such as \xco\ and \cyo, and to the dust disk size, which is measured from the more optically thin millimeter continuum emission originating by millimeter sized dust grains. This optical depth difference between the two tracers creates an inherent observed gas-to-dust size difference, which is difficult to disentangle from real physical differences (e.g. \citealt{Dutrey1998,GuilloteauDutrey1998,Hughes2008,Panic2008,Andrews2009,Facchini2017,Trapman2019}). 

Using detailed thermochemical models that include dust evolution (see \citealt{Facchini2017}), \cite{Trapman2019} showed that a gas-to-dust size ratio larger than four is a ``smoking gun'' for pebble drift. For smaller gas-to-dust size ratios the contribution of optical depth to the ratio must be modeled before we can identify radial drift (see e.g. \citealt{Trapman2020b}).

The characteristics of the observations also play an important role. If the sensitivity is low the faint emission in the outer disk can be lost in the noise, leading to an underestimation of the observed disk size. The dust disk size is often more affected than the gas, as the dust emission decreases more steeply with radius than the optically thick CO emission (e.g. \citealt{Hughes2008,Tazzari2017,Sierra2021,Ilee2022}).
The resolution of the observations can also be a limiting factor, especially if it is comparable to the size of the disk. In the extreme case where both the gas and dust disk are unresolved the size ratio will be one, but more commonly the smaller dust disk will be unresolved and the more extended gas disk could still be, at least marginally, resolved. If uncorrected, this would lead to an underestimation of the gas-to-dust size ratio. This is especially an issue given the fact that recent survey have revealed a large fraction of compact (dust) disks (e.g. \citealt{Hendler2020,Sanchis2021,Miotello2021}).

By fitting a model to the emission, we can try to account for the effect of resolution and limiting the impact of sensitivity. Some care has to be taken that the fitted model can represent the data so as to not bias the results, but in practice this is often possible.
The fit can be done in the uv-plane, i.e., directly fitting the observed visibilities. This approach has seen widespread use for the continuum emission using a range of model intensity profiles, including simple functions such as Gaussian or Nuker profile (e.g. \citealt{Tripathi2017,Andrews2018b,Hendler2020,Sanchis2021}), physical models (e.g \citealt{Tazzari2017,Tazzari2021}) or non-parametric models (e.g. \citealt{jennings2020}).

Fitting the gas emission in a similar fashion is more complex. In order to maximize signal-to-noise, the fitted model should include the Keplerian rotation pattern of the gas, which introduces a dependence between frequency and position. Furthermore, the gas emission is often coming from an elevated layer of the disk, meaning the model cannot be a geometrically thin disk (e.g. \citealt{Dutrey2017,Pinte2018,Izquierdo2021,Law2021bMAPS,Law2022,Paneque-Carreno2023}). Finally, the gas emission, while generally brighter than the continuum, only comes from a narrow velocity range which means the signal-to-noise cannot be increased by using a larger bandwidth.
Visibility modeling of the gas emission is therefore possible (e.g. \citealt{Pietu2007,Guilloteau2011,KurtovicPinilla2024}), but mostly limited to individual bright disks.

For larger samples of disks it is therefore more common to fit the gas emission in the image plane. \citealt{Sanchis2021} fitted the \co\ 2-1 emission of 42 disks in the Lupus star-forming region. For the majority of disks (32/42) they did so using the CASA task \texttt{imfit}\footnote{\url{https://casadocs.readthedocs.io/en/latest/api/tt/casatasks.analysis.imfit.html}} \citep{CASA2022}, which assumes that the observed emission is described by 2D Gaussian that was convolved with the clean beam of the observations. For the remaining ten disks where this approach left significant residuals they instead fit the azimuthally averaged \co\ 2-1 intensity profile with a Nuker profile (characterized by inner and outer power-laws merged around a break radius).

Here we implement the approach of \citealt{Sanchis2021} using the \texttt{astropy.modeling} python package to fit a large diversity of 2D model intensity profiles. We use this to fit clean-beam convolved 2D Nuker and S\'ersic intensity profiles (an exponential profile commonly used to describe intensity profiles of galaxies (e.g. \citealt{Sersic1963,GrahanDriver2005}), to the \co\ 2-1 moment zero maps of twenty disks in Lupus and Upper Sco from the AGE-PRO ALMA large program (see \citealt{AGEPRO_I_overview}). 
The AGE-PRO sample consists of thirty disks around M3-K6 type stars selected from young disks in 
Ophiuchus (embedded objects, $\lesssim$ 1 Myr; e.g. \citealt{Evans2009}), intermediate age disks in Lupus ($\sim1-3$ Myr; e.g. \citealt{comeron2008,Galli2020,AGEPRO_III_Lupus}), and old disks in Upper Sco ($\sim2-6$ Myr; e.g. \citealt{Pecaut2012,Briceno-MoralesChaname2023,Ratzenbock2023,AGEPRO_IV_UpperSco,AGEPRO_VIII_ext_phot_evap}). 
The ten disks in Ophiuchus are not included in our analysis as their \co\ emission is heavily contaminated by extended structures, such as infalling envelopes, outflows, and the molecular cloud (see \citealt{AGEPRO_II_Ophiuchus}).
From the unconvolved best fit models we measure \rgas, the radius that encloses 90\% of the \co\ 2-1 emission, which we compare with \rdust, the radius that encloses 90\% of the 1.3 millimeter continuum emission, from the visibility fits of the dust continuum emission from \cite{AGEPRO_X_substructures}. 

This paper is structured as follows: In Section \ref{sec: methods} we describe our methods and choice of model profile, and discuss how we obtained disk geometries for disks with unresolved continuum emission. In Section \ref{sec: results} we compare the gas disk sizes of our unconvolved best fit models to those measured directly from the images. We then compute gas-to-dust size ratios and 
discuss our results in the context of dust evolution models.
In Section \ref{sec: discussion}, we discuss possible explanations and caveats of this trend and we examine the residuals left in the \co\ moment zero map after subtracting the best fit model. 
Our conclusions are summarized in Section \ref{sec: conclusions}.

%%% last updated: 03 June 2024
\begin{table*}[htb]
\centering
\caption{\label{tab: best fit Nuker profiles} Model disk sizes and best fit Nuker profile parameters }
\def\arraystretch{1.2}%  1 is the default
\begin{tabular*}{0.95\textwidth}{l|ccc|ccccccccc}
\hline\hline
name  &  $R^{\rm model}_{\rm CO,\ 90\%}$ & & &  $A$ & $x_0$ & $y_0$ & $R_{b}$ & $\gamma$ & $\alpha$ & $\beta$ & PA & inc \\
      &  (median) &  (16$^{\rm th}$) & (84$^{\rm th}$) & [Jy/arcsec$^2$ &  &  &  &  &  &  &  &  \\
      &  [arcsec] &  [arcsec] & [arcsec] &  m/s] & [mas] & [mas] & [arcsec] &  &  &  & [deg] & [deg] \\
\hline
Lupus 1  &  1.06 &  1.05 & 1.07 & 332.70 & -40 & 7 & 2.78 & 0.19 & 2.32 & 40.0 & 288.2 & 64.6\\
Lupus 3  &  0.68 &  0.62 & 0.70 & 26.97 & -26 & 32 & 1.38 & 0.90 & 3.69 & 40.0 & 198.4 & 55.2\\
Lupus 4  &  0.19 &  0.04 & 0.40 & 18.42 & -16 & 8 & 0.52 & 1.43 & 10.65 & 20.0 & 199.9 & 36.9\\
Lupus 5  &  0.33 &  0.26 & 0.45 & 148.51 & -32 & 1 & 0.16 & 0.61 & 100.00 & 4.0 & 108.0 & 0.0\\
Lupus 7  &  0.45 &  0.27 & 0.50 & 17.33 & -32 & 5 & 1.96 & 0.96 & 2.06 & 40.0 & 335.8 & 59.0\\
Lupus 8  &  0.48 &  0.47 & 0.49 & 252.72 & -6 & 7 & 0.19 & 0.99 & 100.00 & 3.2 & 260.0 & 67.8\\
Lupus 9  &  - &  - & $<0.09$ & 100.16 & 0 & 0 & 3.41 & 2.00 & 0.50 & 40.0 & 65.0 & 56.8\\
Lupus 10  &  5.38 &  5.03 & 5.71 & 6.16 & 0 & 0 & 20.00 & 0.51 & 1.79 & 40.0 & 110.0 & 50.0\\
Upper Sco 1  &  1.27 &  1.25 & 1.30 & 55.11 & -60 & 23 & 1.44 & 0.49 & 4.71 & 13.7 & 45.0 & 36.0\\
Upper Sco 2  &  0.37 &  0.20 & 0.45 & 1.51 & -49 & -16 & 1.72 & 1.76 & 0.50 & 4.3 & 49.4 & 56.4\\
Upper Sco 3  &  0.24 &  0.23 & 0.27 & 474.23 & -17 & -7 & 0.24 & -0.31 & 100.00 & 30.0 & 77.7 & 57.5\\
Upper Sco 4  &  0.34 &  0.33 & 0.35 & 109.36 & -35 & 2 & 0.59 & 1.00 & 4.00 & 40.0 & 249.0 & 76.0\\
Upper Sco 5  &  0.21 &  0.20 & 0.22 & 70.16 & 6 & 2 & 0.23 & 1.06 & 100.00 & 40.0 & 132.7 & 0.0\\
Upper Sco 6  &  1.01 &  0.94 & 1.10 & 2199.81 & -16 & -1 & 0.23 & -3.00 & 1.47 & 4.9 & 252.9 & 51.4\\
Upper Sco 7  &  1.05 &  1.04 & 1.06 & 28.54 & 45 & -42 & 1.30 & 0.80 & 13.55 & 40.0 & 189.2 & 38.8\\
Upper Sco 8  &  1.04 &  1.02 & 1.05 & 136.74 & -14 & 8 & 2.37 & 0.32 & 2.29 & 40.0 & 16.0 & 16.2\\
Upper Sco 9  &  1.32 &  1.25 & 1.36 & 145000.00 & -10 & 9 & 12.76 & -0.83 & 0.71 & 40.0 & 303.0 & 50.1\\
Upper Sco 10  &  0.55 &  0.52 & 0.56 & 65.81 & -16 & 5 & 0.63 & 1.02 & 6.41 & 13.0 & 334.6 & 49.3\\
\hline\hline
\vspace{0.15cm}
& & & & \multicolumn{9}{c}{best fit S\'ersic profile parameters} \\
\hline
 name & $R^{\rm model}_{\rm CO,\ 90\%}$ & & &  $I_e$ & $x_0$ & $y_0$ & $r_{e}$ & $n$ & $ellip$ & $\theta$ & & \\
      &  (median) &  (16$^{\rm th}$) & (84$^{\rm th}$) & [Jy/arcsec$^2$ &  &  &  &  &  &  &  &  \\
      &  [arcsec] &  [arcsec] & [arcsec] &  m/s] & [mas] & [mas] & [arcsec] &  &  & [rad] &  &  \\
\hline
Lupus 2  &  2.01 &  2.00 & 2.02 & 87.45 & 0 & 0 & 0.93 & 0.94 & 0.24 & 4.1 & & \\
Lupus 6  &  0.33 &  0.31 & 0.37 & 121.92 & 0 & 0 & 0.24 & 0.03 & 0.33 & -0.2 & & \\
\hline\hline
\end{tabular*}
\begin{minipage}{0.91\textwidth}
\vspace{0.1cm}
{\footnotesize{\textbf{Notes:} The amplitude $A$ is related to $I_b$ as $A = 2^{\frac{\beta - \gamma}{\alpha}} I_b$. 
For Lupus 2 and Lupus 6 a Nuker profile does not provide a good description of the observed intensity and we use a S\'ersic profile instead. The parameters of the best fit S\'ersic profile for these two sources are listed at the bottom of the table. For Lupus 9, we take the 84$^{\rm th}$ value as an upper limit of the gas-disk radii as this source appears marginally resolved in \co\ emission. Alternative names (indexed in SIMBAD/CDS) for these sources are presented in Table \ref{tab: keplerian mask parameters}. Host stellar masses and other stellar parameters are presented in \citet{AGEPRO_I_overview}, while the derived AGE-PRO disk masses are presented in \citet{AGEPRO_V_gas_masses}}.
}
\end{minipage}
\end{table*}

%-------------------------------------------------------------------
\section{Methods}
\label{sec: methods}

The AGE-PRO ALMA Large program (\citealt{AGEPRO_I_overview}, 2021.1.00128.L) has an image beam size of $\sim0.15-0.30$ arcsec, the average continuum sensitivity of the survey is 0.025 mJy/beam rms, and the achieved peak \co\ SNR is in the range 15-150. More information on the data properties of individual objects can be found in the AGE-PRO papers of the Lupus (\citealp{AGEPRO_III_Lupus}) and UppSco (\citealp{AGEPRO_IV_UpperSco}) regions. Host stellar masses and other stellar parameters of the AGE-PRO sample are presented in \citet{AGEPRO_I_overview}, while the derived AGE-PRO disk masses are presented in \citet{AGEPRO_V_gas_masses}.

\subsection{Fitting \co\ 2-1 emission in the image plane }
\label{sec: fitting moment zero}

To measure the gas disk sizes we follow a modified version of the method\footnote{A full implementation of this method can be found here: 
\url{https://zenodo.org/records/15360478}}
outlined in \cite{Sanchis2021}:

\begin{enumerate}
    \item We obtained the calibrated \co\ 2-1 spectral cubes of the ten AGE-PRO sources in Lupus from \cite{AGEPRO_III_Lupus} and the ten AGE-PRO sources in Upper Sco from \cite{AGEPRO_IV_UpperSco}. 
    
    \item For each source we construct a Keplerian mask based on the stellar mass, disk orientation and height of the emitting layer using \texttt{keplerian\_mask} function in the \texttt{gofish}\footnote{\url{https://github.com/richteague/gofish}} python package \citep{GoFish}. The Keplerian mask parameters are summarized in Table \ref{tab: keplerian mask parameters} in the appendix. We note here that the parameters are chosen such that the mask conservatively encloses all of the line emission of each source and they should therefore not be interpreted as disk parameters.
    The spectral cubes and Keplerian masks are then combined to make integrated intensity (moment zero) maps using the \texttt{bettermoments}\footnote{\url{https://github.com/richteague/bettermoments}} python package \citep{TeagueForeman-Mackey2018,Teague2019}. 
    This also yields noise maps that describe how the uncertainty in the moment zero map varies spatially after inclusion of a Keplerian mask
    
    \item Using the \texttt{astropy.modeling}\footnote{\url{https://docs.astropy.org/en/stable/modeling/index.html}} python package we fit these moment zero maps using three different model intensity profiles, Gaussian, Sersic, and Nuker\footnote{A Nuker profile is currently not available among the \texttt{astropy.modeling.Fittable2Dmodels}, but an implementation can be obtained here: 
    \url{https://zenodo.org/records/15360478}}
    (see Appendix \ref{app: profile math}). 
    A 2D version of the chosen profile is convolved with the clean beam (using \texttt{astropy.convolution.convolve\_models}) before it is fitted to the moment zero map using \texttt{astropy.modeling.LMLSQFitter}, which uses the Levenberg-Marquardt algorithm as implemented by \texttt{scipy.optimize.least\_squares} \citep{More2006}. Disk geometries (inclinations and position angles) are fitted as free parameters, with the caveat for the interpretation that the CO is not coming from the midplane (see \citealp{AGEPRO_X_substructures} for continuum derivations of AGE-PRO disk geometries).

    \item For the weights we adopt $1/\left(\sigma_{\rm bm} \times N_{\rm pixel-per-beam}  \right)$, where $\sigma_{\rm bm}$ is the noise map obtained with \texttt{bettermoments} and $N_{\rm pixel-per-beam}$ is the number of pixels per clean beam. The latter factor is include to account of the fact that the noise of pixels in the same beam are correlated. The least squares fitters fits on a pixel-by-pixel basis and assumes that the noise of each pixel is independent. In doing so it will underestimate the noise by approximately a factor $\sqrt{N_{\rm pixel-per-beam}}$ compared to fitting on a beam-by-beam basis. We note however that this is only an approximate correction for the correlated noise in interferometric images (see, e.g., \citealt{Davis2017} for a more correct implementation). 

    \item After the fit has converged we take the unconvolved, best-fitting model and measure the radius (in the disk coordinate frame) that encloses 90\% of its total flux using a curve-of-growth method $(R^{\rm model}_{\rm CO,\ 90\%})$. This is identical to the method used in the accompanying AGE-PRO papers to measure the observed gas disks size in the image plane. 

    \item To compute the uncertainties on $R^{\rm model}_{\rm CO,\ 90\%}$ we use a Monte Carlo approach similar to the one used in \cite{Sanchis2021}.
    We draw sets of model parameters from a multivariate normal distribution with means equal to parameters of the best fit model and the covariance matrix computed by \texttt{scipy.optimize.least\_squares} as part of the fit. 
    For each set of parameters we construct a model profile and measure its $R^{\rm model,i}_{\rm CO,\ 90\%}$ in identical fashion as for the best fitting model. We take the 16$^{\rm th}$ and 84$^{\rm th}$ as the lower and upper uncertainty on $R^{\rm model}_{\rm CO,\ 90\%}$. 
\end{enumerate}

\subsection{The choice of model intensity profile}
\label{sec: the choice of model intensity profile}

\begin{figure}[!ht]
    \centering
    \includegraphics[width=\columnwidth]{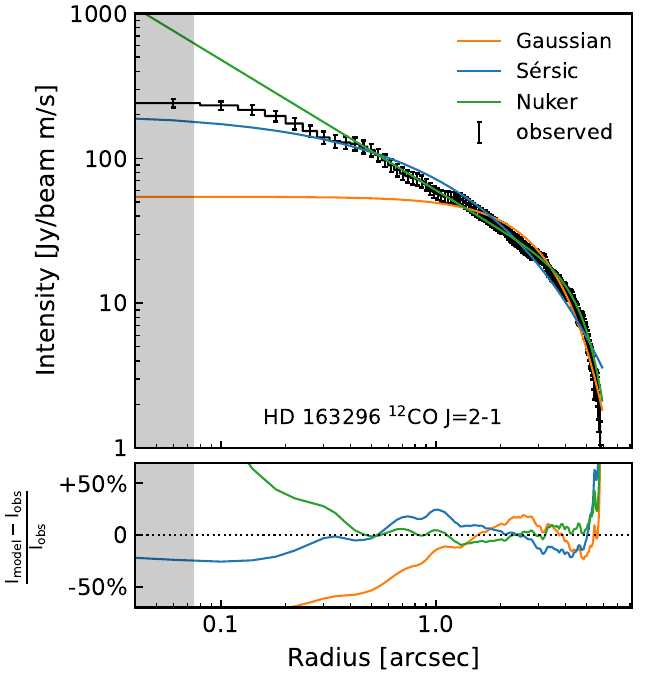}
    \caption{\label{fig: HD163296 example profile} Comparison between the observed \co\ 2-1 radial intensity profile (in black) of HD 163296 \citep{Law2021bMAPS} and the three model intensity profiles discussed in this work: Gaussian (orange), S\'ersic (blue), and Nuker (green). Bottom panel shows the fractional difference between the observed and model intensities. 
    Note that the focus here is on fitting the outer disk as it produces most of the total flux and therefore has the largest effect on the measured gas disk size.}
\end{figure}

When fitting the observations, one of the important choices is which model we use. If the chosen model does not accurately describe the intensity distribution of the observed disk the best fit model, and any disk size derived from it, would be biased. In this work we consider three different model intensity profiles: a Gaussian, a S\'ersic profile and a Nuker profile (see Appendix \ref{app: profile math} for analytic prescriptions of each). 

To get a better understanding of how these profiles compare to each other and to the disk \co\ intensity profile, in Figure \ref{fig: HD163296 example profile} we compare them to a high resolution ($\sim0\farcs15$) \co\ 2-1 intensity profile of the disk around HD 163296 (see \citealt{Law2021bMAPS}). It is clear that the Gaussian profile is not a good description of observed intensity. It underproduces the flux in the inner $\sim1\farcs0$ but describes the intensity in the outer disk reasonably well. Inside $0\farcs4$ the S\'ersic profile lies close to the observations, where outside $0\farcs4$ the Nuker profile provides a better description of the observations. The latter is perhaps unsurprising, as the Nuker profile matches well with the expected CO intensity profile. The optically thick CO emission follows the temperature profile in the CO emitting layer, which is usually well described by a powerlaw, followed by a drop when the CO column density drops below the CO self-shielding threshold (e.g. \citealt{Trapman2023}). 

We should note here that since our definition of the gas disk size is linked to the total flux, reproducing the outer disk intensity is more important than reproducing the inner disk intensity, as the former with its larger area dominates the total flux. In this work we therefore opt to fit a Nuker profile to 18 out of the 20 AGE-PRO disks. However, note that Figure \ref{fig: comparing Nuker and Sersic} in Appendix \ref{app: comparing Nuker and Sersic} shows that Nuker and S\'ersic profiles yield very similar gas disk sizes for almost all of the sources studied in this work. The Nuker profile is given by: 

\begin{equation}
\label{eq: Nuker profile}
    I(x,y)  = 2^{\frac{\beta - \gamma}{\alpha}} I_b \left(\frac{R}{R_b}\right)^{-\gamma} \left[1 + \left(\frac{R}{R_b}\right)^{\alpha}\right]^{\frac{\gamma-\beta}{\alpha}}.
\end{equation}
Here $I_b$ is the intensity at the break radius $R_b$, $\gamma$ and $\beta$ are, respectively, the powerlaw slopes of the powerlaws for radii inside and outside $R_b$, respectively. The parameter $\alpha$ describes the transition between these two powerlaws, with $\alpha \rightarrow \infty$ yielding a sharp, broken powerlaw and small values of $\alpha$ providing a smooth transition between the two powerlaws. It is worth mentioning that the Nuker profile only provides a finite luminosity if $\alpha > 0$, $\beta >0$ and $\gamma <2$ (e.g. \citealt{Baes2020}).  

The best fit parameters are summarized in Table \ref{tab: best fit Nuker profiles}. As a test we also fitted each moment zero map with a S\'ersic profile, finding very similar residuals and \rgasmodel\ to what was found for the best fitting Nuker profiles (see Appendix \ref{app: comparing Nuker and Sersic}). The exceptions are Lupus 2 (GW Lup/Sz 71) and Lupus 6, where the best fitting S\'ersic profile provides a much better description of the observations than the best fitting Nuker profile. This could be due to the presence of strong cloud contamination, which we discuss in the next section, although the fact that Lupus 6 could potentially be a triple system (see \citealt{AGEPRO_III_Lupus}) likely also plays a role. For these disks we therefore use the best fitting S\'ersic profile for our analysis instead. 

\subsection{Cloud contamination and gas size measurements}
\label{sec: cloud contamination}

Several sources in the sample, especially those in Lupus, suffer from cloud contamination. In most cases this is in the form of cloud absorption of the \co\ emission around the systemic velocity, but in some cases \co\ cloud emission can also be seen in the channel maps (see \citealt{AGEPRO_III_Lupus}). Both will affect the measurement of \rgas. Cloud absorption predominantly removes emission at the velocity of the cloud, which is close to the systemic velocity of the disk. In a Keplerian rotating disk, emission at velocities close to the systemic velocity originate from the outer disk. By removing emission from the outer disk, cloud absorption will therefore reduce the radius that encloses 90\% of the total flux in the observed image (see also Appendix B in \citealt{Long2022}). Note, however, that in most cases cloud absorption only removes emission on one side of the disk, meaning that \rgas\ can be measured from the uncontaminated half.  
Cloud emission can instead increase the size of the disk by adding a significant amount of flux at large separations from the disk. Using a Keplerian mask will in most cases mask out this emission as the cloud does not follow Keplerian rotation. 

\begin{figure*}[htb]
    \centering
    \includegraphics[width=0.9\textwidth]{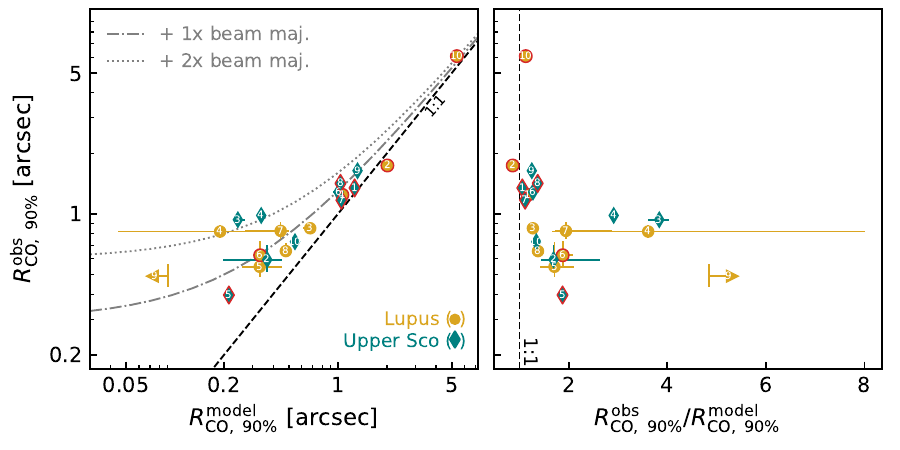}
    \caption{\label{fig: comparing model and observed size} Comparison of the observed $(\rgasobs)$ and model $(\rgasmodel)$ gas disk size. Lupus and Upper Sco disks are shown as yellow circles and teal diamonds, respectively, with the numbers corresponding to the source IDs in Table \ref{tab: best fit Nuker profiles} (see \citealt{AGEPRO_I_overview} for details). Sources with significant residuals after subtracting the best fit model are denoted with a red outline.
    Gray dashed-dotted and dotted lines shows differences between the observed and model gas disk size of one and two times the typical beam major axis $(\sim0\farcs3)$ of the AGE-PRO \co\ observations. }
\end{figure*}

Based on \cite{AGEPRO_III_Lupus}, cloud contamination is (most) present for Lupus 2, 6, 8, and 10. 
Similarly, some cloud contamination can be seen for Upper Sco 6 \citep{AGEPRO_IV_UpperSco}. 
To minimize its effect on both the fit of the gas emission and the resulting measurement of \rgasmodel, while running the fit we mask out the position angles of the moment zero map where cloud absorption can be seen in the channel maps.
Most of the cloud emission is excluded by the Keplerian masks. We carry out a visual inspection of the channel and moment zero maps. In cases where there is significant emission the moment map that is unrelated to the disk we also mask out everything outside a maximum radius (measured in the disk coordinate frame).
The masked regions for each source are shown in Figures \ref{fig: CO moment fits Lupus 1 to 3},\ref{fig: CO moment fits Lupus 4 to 7}, and \ref{fig: CO moment fits Lupus 8 to 10} in Appendix \ref{app: fit results}.

%-------------------------------------------------------------------
\section{Results}
\label{sec: results}

\subsection{Fitted gas disk sizes}
\label{sec: fitted gas disk sizes}

Figure \ref{fig: comparing model and observed size} shows a comparison between the gas disk size measured from the unconvolved fitted model (\rgasmodel) and the size measured directly from the images (\rgasobs), the latter is obtained from \cite{AGEPRO_III_Lupus} and \cite{AGEPRO_IV_UpperSco} for disks in Lupus and Upper Sco, respectively. Eight disks have significant ($\geq5\sigma$) residuals left after subtracting the convolved best fit model (see Sect. \ref{sec: residuals}). In Figure \ref{fig: comparing model and observed size} these sources are denoted with a red outline.

For disks with $\rgasobs\gtrsim 1\farcs0$ the observed size and the unconvolved model size are nearly identical.  These larger disks are much larger than the beam and the effect of beam convolution on \rgasobs\ is therefore small. For an analytical example of this, consider a disk with a Gaussian intensity profile  with a size $\sigma_{\rm disk}$ that is convolved with a Gaussian beam with size $\sigma_{\rm beam}$. 
The size of the convolved Gaussian is given by 
\begin{equation}
\sigma_{\rm disk,\ conv} = \sqrt{\sigma_{\rm disk}^2 + \sigma_{\rm beam}^2}  = \sqrt{1 + \tfrac{\sigma_{\rm beam}^2}{\sigma_{\rm disk}^2}}\times \sigma_{\rm disk}    
\end{equation}
which converges to $\sigma_{\rm disk,\ conv} \approx \sigma_{\rm disk}$ for $\sigma_{\rm disk} \gg \sigma_{\rm beam}$. 

The reverse is of course true if $\sigma_{\rm beam} \gtrsim \sigma_{\rm true}$, in which case $\sigma_{\rm disk,\ conv} \approx \sigma_{\rm beam}$. The smaller disks in the sample reflect this, with differences between \rgasmodel\ and \rgasobs\ increasing as \rgasobs\ decreases. The right panel of Figure \ref{fig: comparing model and observed size} shows that the difference can be up to twice the major axis of the typical beam in AGE-PRO $(\sim 0\farcs3)$. 

Lupus 2 stands out by the fact that its model size is larger than the size from the convolved image. This disk suffers from substantial cloud absorption, which in particular removes emission at large radii for the northeastern half of the disk. As discussed in Sect. \ref{sec: cloud contamination}, this reduces \rgas\ if measured from the full moment zero map, as is the case for \rgasobs. As the model was fit to the uncontaminated half of the disk (see Figure \ref{fig: CO moment fits Lupus 1 to 3}), \rgasmodel\ is not affected by the cloud and we thus find a larger gas disk size.   

\begin{figure}
    \centering
    \includegraphics[width=0.9\columnwidth]{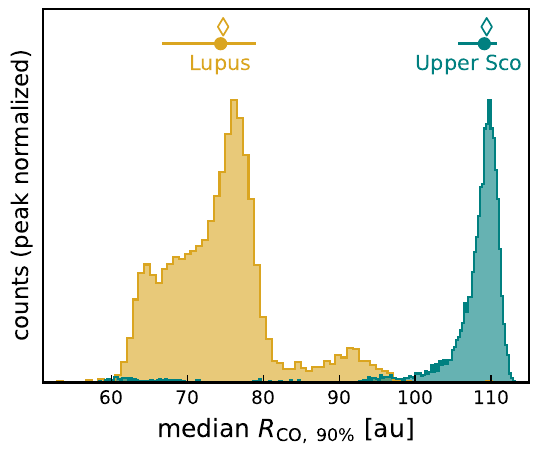}
    \caption{\label{fig: median gas disk sizes} Distributions of the median gas disk size for the complete bootstrapped population of} disks in Lupus (yellow) and Upper Sco (green). Circles above the distribution show the median of the distribution and the vertical lines stretches from its 16$^{\rm th}$ to its 84$^{\rm th}$ quantile. Note here that the uncertainty on the median only includes the propagated observational uncertainties and not the effect of our limited sample size. Open diamonds show the median values from \cite{AGEPRO_I_overview} for AGE-PRO only sources. 
\end{figure}

In the context of disk evolution it is also interesting to compare the median gas disk sizes in the two regions. We compute the medians using a Monte Carlo method, where for each disk we take a random disk size from its posterior distribution and measure the median for this collection of sizes. This process is repeated 1000 times to construct distributions of median disk sizes for disks in Lupus and Upper Sco, which are shown in Figure \ref{fig: median gas disk sizes}.

\begin{figure}[htb]
    \centering
    \begin{minipage}{\columnwidth}
    \includegraphics[width=\columnwidth,clip,trim={0 1.2cm 0 0}]{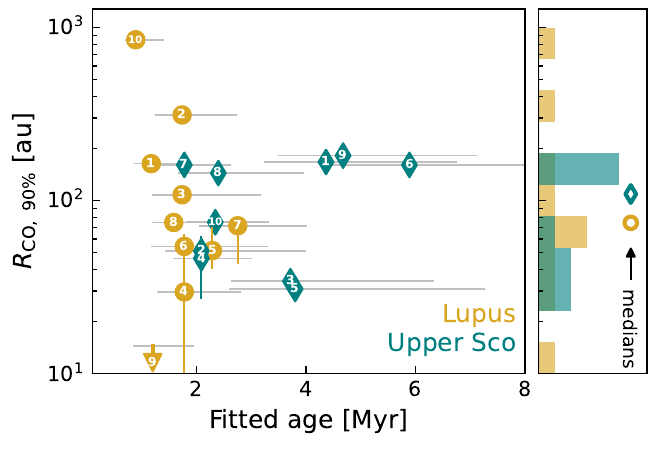}    
    \end{minipage}
    \begin{minipage}{\columnwidth}
    \includegraphics[width=\columnwidth]{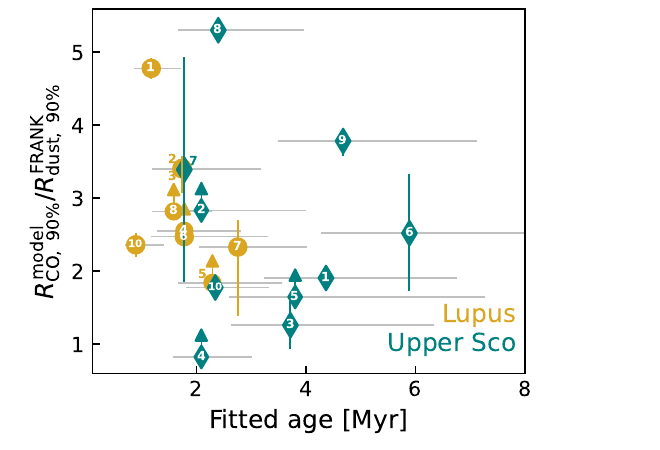}    
    \end{minipage} 
    \caption{\label{fig: sizes versus age} Gas disk size (top) and gas-to-dust size ratio (bottom) of the Lupus and Upper Sco disks versus isochronal ages of individual sources \citep{AGEPRO_III_Lupus,AGEPRO_IV_UpperSco}. The top right panel shows histograms of the gas disk size, with the median size of each region marked with an open symbol (see also Figure \ref{fig: median gas disk sizes}).}
\end{figure}

We find a smaller median disk size of $\rgasmodel = 74.5_{-7.4}^{+4.4}$ au for the younger disks in Lupus compared to a median size of $\rgasmodel = 109.1_{-3.3}^{+1.5}$ au for the older disks in Upper Sco. At face value this suggests that disks grow over time, in line with the expectation of viscous disk evolution (e.g. \citealt{LyndenBellPringle1974,Pringle1981,Trapman2020}). However, as discussed in e.g. \cite{AGEPRO_IV_UpperSco} there is overlap in age between the Lupus and Upper Sco sources. The top panel of Figure \ref{fig: sizes versus age} shows gas disk sizes versus isochronal ages of individual sources (see \citealt{AGEPRO_III_Lupus,AGEPRO_IV_UpperSco} for details on the ages). There is no clear increase in disk size with individual age, but instead we seem to be seeing a narrowing of the gas size distribution. A likely cause for this is survivorship bias, i.e., the fact that disk evolution, both viscous and wind-driven, will disperse smaller disks faster than larger disks (e.g. \citealt{AGEPRO_VII_diskpop}). 

For a more detailed analysis of the gas disk sizes in the context of disk evolution using disk population synthesis to account for survivorship bias we refer the reader to \cite{AGEPRO_VII_diskpop}.

\subsection{Gas versus dust disk sizes}
\label{sec: gas vs dust disk size}

\begin{figure*}[htb]
    \centering
    \includegraphics[width=\textwidth]{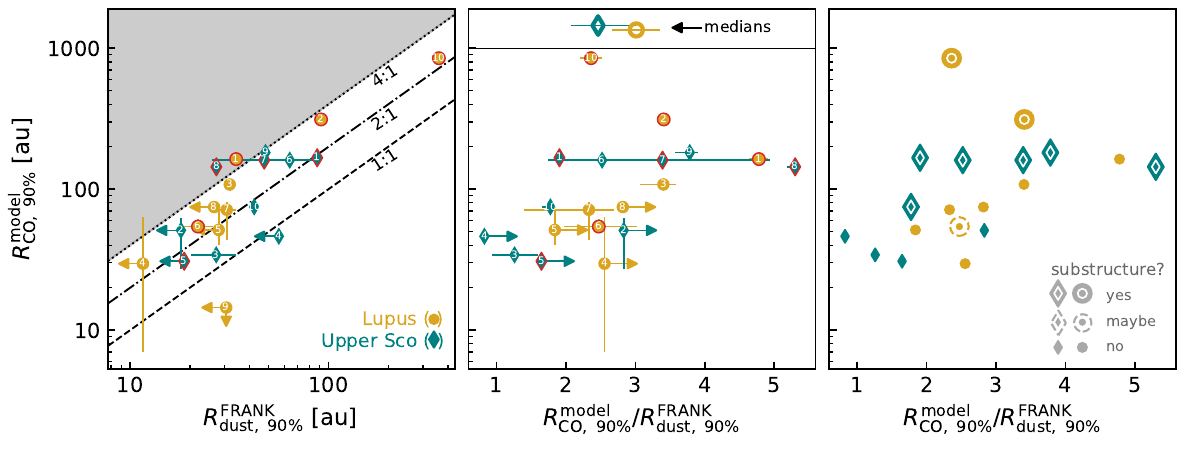}
    \caption{\label{fig: gas vs dust disk size} Comparison of the gas disk size (\rgasmodel), obtained from the unconvolved model, and the dust disk size (\rdustfrank), obtained from fitting in the continuum visibilities using FRANK (see \citealt{AGEPRO_X_substructures} for details).
    Lupus and Upper Sco disks are shown as yellow circles and green diamonds, respectively, with the numbers corresponding to the source IDs in Table \ref{tab: best fit Nuker profiles} (see \citealt{AGEPRO_I_overview} for details).
    The black lines in the left panel show fixed gas-to-dust size ratios, with $\gtrsim 4$ being a clear sign of radial drift (e.g. \citealt{Trapman2019}) while ratios $\lesssim4$ could be explained by optical depth effects, although they do not rule out radial drift.
    The right panel is a repeat of the middle panel, but shows which sources have dust substructures according to the analysis in \cite{AGEPRO_X_substructures}.}
\end{figure*}

The difference between the gas and dust disk size is often considered a measure of how much millimeter dust grains have drifted inward (e.g. \citealt{Andrews2011,Andrews2016,deGregorioMonsalvo2013,Pietu2014,Cleeves2016,Toci2021}) although optical depth differences between the optically thin continuum and optically thick \co\ emission also play an important role (e.g. \citealt{Hughes2008,Panic2008,Trapman2019}).  

We combine the gas disk sizes from this work with the dust disk sizes (\rdustfrank) from \cite{AGEPRO_X_substructures}, who fitted the 1.3 millimeter continuum visibilities of the AGE-PRO sources using FRANK (see \citealt{jennings2020}). 
Figure \ref{fig: gas vs dust disk size} compares the gas and dust disk sizes of the twenty sources discussed in this work. We find gas-to-dust disk size ratios $(\rgasmodel/\rdustfrank)$ between $\sim1$ and $\sim5.5$, with a median of $\langle\rgasmodel/\rdustfrank\rangle = 2.78^{+0.37}_{-0.32}$, which is consistent with previous studies (e.g. \citealt{ansdell2018,Kurtovic2021,Sanchis2021,Long2022}). We note there are 4 Lupus and 3 Upper Sco sources with \rdustfrank upper limits because they are only marginally resolved in dust continuum emission (\citealp{AGEPRO_X_substructures}).

Looking now at the two regions separately, we find that a median higher ratio for the younger disks in Lupus ($\langle\rgasmodel/\rdustfrank\rangle_{\rm Lupus} = 3.02^{+0.33}_{-0.33}$) compared to $\langle\rgasmodel/\rdustfrank\rangle_{\rm Upper\ Sco} = 2.78^{+0.53}_{-0.38}$ for the older disks in Upper Sco. Here the uncertainties on the median are computed using a Monte Carlo simulation where we draw a random \rgasmodel/\rdustfrank\ for each source from its uncertainty and compute the median for this sample. The provided values are the median and 16$^{\rm th}$ and 84$^{\rm th}$ quantile of posterior distribution of median \rgas/\rdust. Within these uncertainties the size ratios of the two regions are the same. A similar conclusion can be drawn from the bottom panel of Figure \ref{fig: sizes versus age}, which shows \rgasmodel/\rdustfrank\ as a function of individual isochronal age. No clear trend with age can be seen here, although there might be a hint of a decrease of \rgas/\rdust\ for older sources. 

A decreasing or constant \rgas/\rdust\ with age is the opposite of what is expected from theory. The older disks in Upper Sco should have had more time for grains to drift inward, meaning that we would have expected $\langle\rgas/\rdust\rangle_{\rm Upper Sco} > \langle\rgas/\rdust\rangle_{\rm Lupus}$ (see, e.g. \citealt{Rosotti2019,Toci2021}). 
We will discuss this in more detail in Section \ref{sec: rgas/rdust discussion}.

The rightmost panel of Figure \ref{fig: gas vs dust disk size} again shows the gas-to-dust size ratio versus \rgasmodel, but distinguishes between disks that have substructures, potential substructures or no substructures in their continuum emission based on the visibility modeling in \cite{AGEPRO_X_substructures}. Two features immediately stand out. First, there is no apparent dependence of \rgas/\rdust\ on whether a disk has dust substructures. A similar result was found by \citealt{Long2022} who examined a sample of 44 disks from Taurus, Lupus, and the DSHARP sample \citep{Andrews2018}. The second feature in Figure \ref{fig: gas vs dust disk size} is that substructures are detected in all large ($\rgasmodel \gtrsim 100$ au) disks and rarely so in smaller disks. It is enticing to link this to the formation mechanism of substructures, e.g., 
larger, more massive, disks are more likely to have formed more massive planets that created deeper and wider dust gaps.
But in all likelihood this is a matter of the resolutions of the observations $(\sim0\farcs3)$ not being able to resolve small substructures in compact disks (see \citealt{Bae2023} for a review). Indeed, \cite{AGEPRO_X_substructures} report a typical visibility fitting resolution of $\sim22$ au. If dust disks are typically two times smaller than gas disks, any gas disk smaller than $\sim100$ au would only have $\sim2.5$ of such resolution elements across its dust disk, making it nigh impossible to detect any dust substructures.
High spatial resolution observations are needed to determine if there is indeed a link between the size of the gas disk and the incidence and properties of dust substructures.

\subsection{Fitted Nuker profiles}
\label{sec: fitted Nuker profiles}

\begin{figure*}
    \centering
    \includegraphics[width=\textwidth]{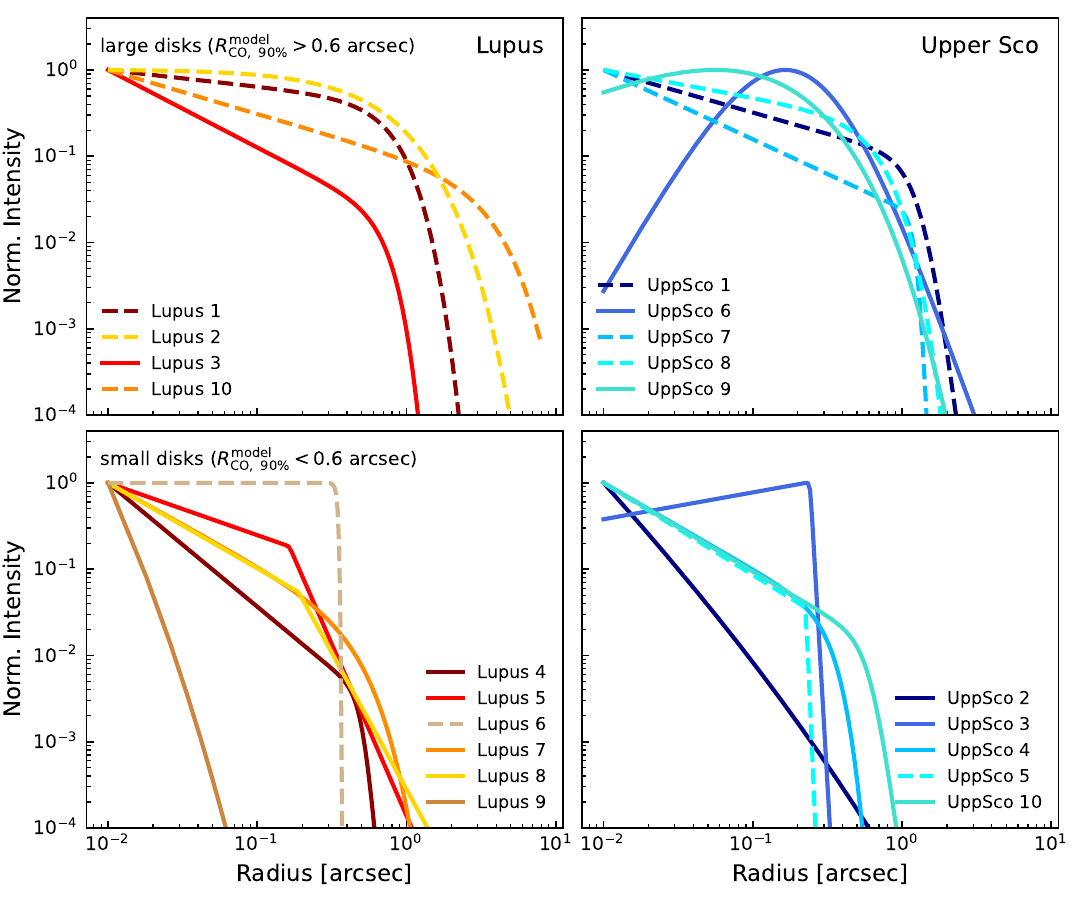}
    \caption{\label{fig: Nuker profiles} Best fit (Nuker), peak-normalized, intensity profiles of the twenty sources examined in this work (see Table \ref{tab: best fit Nuker profiles}). 
    Top and bottom panels show sources with $\rgasmodel < 0\farcs6$ and $\rgasmodel \geq 0\farcs6$, respectively , with sources in Lupus shown on the left and those in Upper Sco on the right.
    Profiles of sources with significant residuals after subtracting the best fit model are shown with a dashed line. The left panel also shows two representative powerlaws, with slopes of $-0.5$ and $-1.0$, as black dashed-dotted lines.}
\end{figure*}

In addition to providing gas disk sizes, fitting the moment zero map also gives an idea of the unconvolved intensity profile of the AGE-PRO disks. Figure \ref{fig: Nuker profiles} shows the best fit intensity profiles, split up into small disks $(\rgasmodel \leq 0\farcs6)$ and large disks $(\rgasmodel > 0\farcs6)$

Starting with the large disks, the intensity profiles follow the general shape expected for \co\ emission (see Section \ref{sec: the choice of model intensity profile}). We find shallow $(\gamma\approx0.3-0.9)$ powerlaw slopes out to a radius of $\sim 1\farcs0$, followed by a steep drop in the \co\ intensity profile. The inner power law slopes are comparable to the expected temperature profile slope for protoplanetary disks ($\sim 0.5-0.75$; e.g. \citealt{ChiangGoldreich1997,DAlessio98}).
It should be noted however that many of these sources, especially Lupus 1, Upper Sco 1, and Upper Sco 8, have significant residuals so the parameters of their best fit model profile should be viewed with caution.

For the small disks we see a much larger diversity of intensity profile shapes, with inner powerlaw slopes as steep as $\gamma\approx1.4$. 
However, in most cases this inner power law lies within a single resolution element $(\sim 0\farcs3)$, meaning that there are only limited constrains from the observations on this part of the intensity profile. 

In a few cases, i.e. Upper Sco 3, 6, and 9, the best fit Nuker profile shows positive inner slope, meaning the model profile has an inner cavity. The cavity radii are small $(\lesssim 0\farcs2)$ and are not directly seen in the moment zero map. The FRANK fit of the Upper Sco 6 also show an inner cavity in the dust with a radius of $\sim0\farcs12$ (see \citealt{AGEPRO_X_substructures}). If the gas cavity seen in the \co\ model is real this would mean that the gas cavity is more extended than the dust cavity, which is the opposite of what is commonly seen in observations (e.g. \citealt{vdMarel2015,vdMarel2016,vdMarel2022}) and hard to explain from theory (e.g. \citealt{Pinilla2012,Pinilla2016,Flock2015,Garate2021,Garate2023}). 
Interestingly, a small number of the FRANK fit realizations for the continuum of Upper Sco 3 show a ring at a similar radial location. However, the continuum intensity profile of Upper Sco 9 shows no inner cavity and several other sources with clear dust cavities do not show a matching cavity in the \co\ emission. 
However, we should stress here that our fitting method is not very sensitive to morphologies on scales smaller than the beam nor are the best fit profiles shown here necessarily unique fits given the uncertainties of the observations. As such, the presence of gas cavities and their cavity radii should be viewed with caution.
High spatial resolution observations are needed to confirm if these disks indeed host small inner cavities.  

% %-------------------------------------------------------------------
\section{Discussion}
\label{sec: discussion}

\subsection{Why does \rgas/\rdust\ not increase with age?}
\label{sec: rgas/rdust discussion}

As mentioned in Section \ref{sec: gas vs dust disk size}, a decreasing or constant \rgas/\rdust\ with age is unexpected from current disk evolution theory. Dust evolution and the inward drift of solids predict that \rgas/\rdust\ increases with time as \rdust\ decreases with time (e.g. \citealt{Rosotti2019,AGEPRO_VI_DustEvolution}). If disks are also evolving viscously the increase of \rgas/\rdust with time is even steeper as now \rgas\ also increases with time (e.g. \citealt{Toci2021}). 

However, our results (Figure \ref{fig: sizes versus age}) contrast with this theoretical expectation. One possible explanation is the presence of one or more dust traps which are halting the inward drift at the trap location (e.g. \citealt{Pinilla2012}).
The resulting concentration of solids will contribute significantly to the total millimeter flux. By measuring the dust disk size as the radius that encloses 90\% of the total flux, \rdust\ is strongly correlated with the location of the outermost dust trap (see \citealt{AGEPRO_VI_DustEvolution} for a detailed analysis and discussion of this; see also e.g. \citealt{Pinilla2020,vdMarelMulders2021,Zormpas2022}). 
If the outermost dust trap has already formed by the age of Lupus ($\sim 1-3$ Myr) \rdust\ would no longer evolve between the ages of Lupus and Upper Sco. The absence of evolution of dust radii with age is indeed observed for the AGE-PRO sample \citep{AGEPRO_X_substructures}. This explanation was also put forward by \cite{Toci2021} to explain the discrepancy between their models and the observations. For the AGE-PRO sources \cite{AGEPRO_VI_DustEvolution} showed that weak or strong dust traps are needed not just to explain the dust disk sizes but also the integrated millimeter fluxes and spectral indices.

\begin{figure*}[thb]
    \centering
    \includegraphics[width=\textwidth]{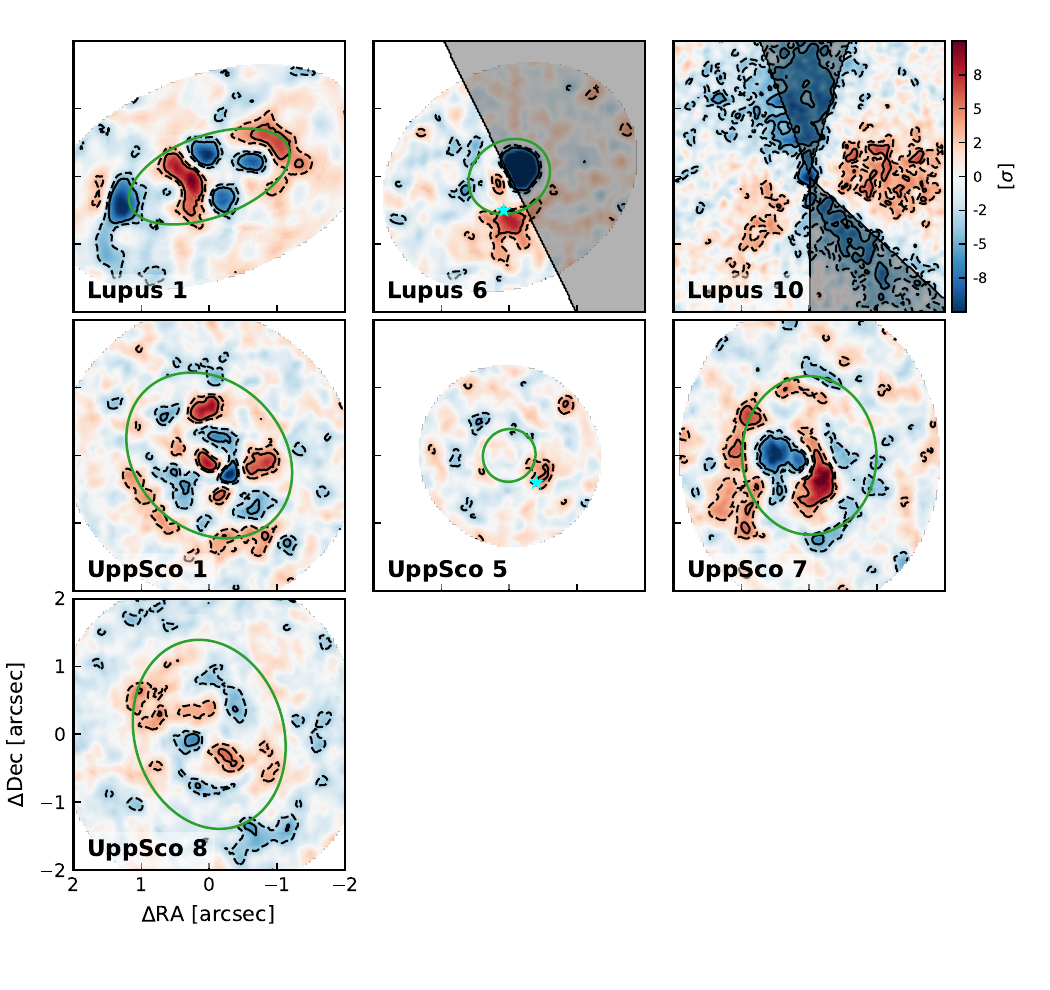}
    \caption{\label{fig: residuals} Residuals after subtracting best fit Nuker/S\'ersic profile for Lupus 1, 6, and 10 and Upper Sco 1, 5, 7, and 8. All residual maps use the same colormap and range. Black solid and dashed contours show $5\sigma$ and $3\sigma$, respectively. The black shaded regions in the images for Lupus 10 and Lupus 6 show what part of the moment zero map was masked during the fit to remove cloud contamination. The green contour shows \rgasmodel\ for each source (see Table \ref{tab: best fit Nuker profiles}). Note for Lupus 10 \rgasmodel\ lies outside the image shown here. 
    Finally, cyan stars in Lupus 6 and Upper Sco 5 mark the locations of $3\sigma$ continuum peaks in the imaged continuum residuals (see \citealt{AGEPRO_X_substructures} for details)}. 
\end{figure*}

In addition to halting the radial drift, external photoevaporation can also help explain the lack of evolution of the gas-to-dust size ratio, or its potential decrease with time, by reducing the gas disk size. 
The disks in Upper Sco are in the proximity of a large numbers of B-type stars and therefore subjected to much higher levels of external irradiation than the disks in Lupus where such stars are mostly absent (e.g. \citealt{Trapman2020,AGEPRO_VIII_ext_phot_evap}). \cite{AGEPRO_VIII_ext_phot_evap} analyzed the effect on external FUV irradiation on \rgas\ and showed that \rgas, and the gas-to-dust size ratio, of the AGE-PRO Upper Sco disks could be significantly reduced by it. 
In particular, they also show that combining external photo-evaporation with dust evolution in a smooth disk cannot explain the observed gas-to-dust size ratios in Upper Sco (see \citealt{AGEPRO_VIII_ext_phot_evap} for details). 

The explanation given here depends strongly on the presence of substructures in the disk capable of halting pebble drift. \cite{AGEPRO_X_substructures} modeled the continuum visibilities of the AGE-PRO sources and find substructures in six Upper Sco and three Lupus disks, with one of the latter being tentative. This suggests that dust traps are present by the age of Upper Sco, but leaves it unclear whether such traps are also already present at the younger age of Lupus. However, the non-detection of substructure does not equal its absence, especially given the low resolution of the observations relative to the typical dust disk size of the Lupus disks (see \citealt{AGEPRO_X_substructures}). Indeed, \cite{AGEPRO_VI_DustEvolution} showed that dust traps are also necessary to match the rate of decrease of the millimeter fluxes and the increase of spectral index with age that is seen in the AGE-PRO sample. These results provide strong indirect evidence for a population of unresolved substructures in the AGE-PRO disks. 

It should also be kept in mind that the comparison between the two regions could be biased by the limited sample size, the method by which the sources were selected and survivorship bias. The sources in both regions were selected to have a range of millimeter luminosities similar to the range of luminosities of the whole class II disk sample in their respective star-forming region (see \citealt{AGEPRO_I_overview} for details). \cite{AGEPRO_X_substructures} showed that the same is true for \rdust, at least within the spectral type range of AGE-PRO.
The situation is different for the gas. The AGE-PRO sources were selected from sources with previous \co\ 2-1 or 3-2 detections, which biases the sample towards larger gas disks. If the AGE-PRO sources are indeed relatively unbiased in their dust disk size, a bias towards larger \rgas\ would mean that the samples are biased towards larger gas-to-dust size ratios. It is unclear how this affects the comparison between the two regions. The overall literature detection rate of \co\ is lower in Upper Sco than in Lupus, but within the M3-K6 spectral type range of AGE-PRO the detection rates are similar (79\% versus 88\%, e.g. \citealt{ansdell2018,Barenfeld2016,UpperSco_followup}, see \citealt{AGEPRO_I_overview}). However, these previous surveys targeted different lines, \co\ 2-1 in Lupus and \co\ 3-2 in Upper Sco, and have different line sensitivities, so a direct comparison of detection rates should be viewed with caution.

\subsection{Residuals after subtracting the best fit model}
\label{sec: residuals}

While in general our models are able to accurately reproduce the observations, for eight sources we find significant (multiple $\geq5\sigma$ peaks) residuals after subtracting our best fit model from the observed \co\ moment zero map (see, e.g., Figure \ref{fig: CO moment fits Lupus 1 to 3} in Appendix \ref{app: fit results}).
Two of these sources are Lupus 2 and 6, which are heavily affected by cloud contamination. For these sources it is hard to determine whether residuals are due to the cloud or due to a mismatch between the model and the observations. For Lupus 2 that is reason to include it from the analysis here, but as we will discuss below, Lupus 6 exhibits residual emission that likely has a different origin than the cloud.

Figure \ref{fig: residuals} shows the residual maps for these seven sources. We find a range of different residual patterns, from highly structured such as the apparent sine wave along the major axis for Lupus 1 and the X-shape for Lupus 10 to the more asymmetric residuals of Upper Sco 1, 7 and 8. The majority of the sources are large disks, with the exception of Upper Sco 5 and Lupus 6, which are among the smallest disks examined in this work. Below we discuss each source individually.

\emph{Lupus 1}: the residuals for this disk follow a sinusoidal pattern along the major axis of the disk. They start negative at the Eastern edge of the disk, followed by positive and negative residuals the East- and West side of the disk respectively before ending with positive residuals at the Western edge of the disk. A potential explanation for this pattern is a East-West asymmetry, i.e., the true slope of the intensity profile is shallower on the East side and steeper on the West side compared to our best fit model that assumed azimuthal symmetry. Interestingly, \cite{Miley2024} fitted a Keplerian model to their higher spatial resolution \co\ 2-1 moment one map and reported a slight but coherent residual on the Western side of the disk, which appears to be consistently redshifted with respect to the Eastern side of the disk. 

If there is indeed a East-West asymmetry, it is unclear what caused it. \cite{Miley2024} rule out disk geometry as the cause for their residual signal and our best fit model also appears to reproduce the overall shape of the \co\ 2-1 integrated intensity. As the optically thick \co\ emission traces the disk temperature, an East-West asymmetry in its emission implies an East-West asymmetry in the temperature profile, at least in the \co\ emitting layer. 

\emph{Lupus 10 (and Upper Sco 8)}: The residuals for Lupus 10 follow a pattern of positive residuals along the major axis of the disk and negative residuals along the minor axis. These are signs of subtracting a geometrically thin disk model from a source where the emission is coming from an elevated emitting layer.
For lines where emission surface is elevated from the midplane the moment zero map resembles an elongated X shape (see, e.g., the \co\ 2-1 moment zero map of HD 163296 in \citealt{Law2021bMAPS}). The two arms on one side of the major axis are the emitting surfaces of the front side of the disk, the two arms on the other side are the emitting surfaces of the back side. 
When fitting an azimuthally symmetric model intensity profile to such a moment zero map, it will underpredict the emission of the front side emitting surface, leading to positive residuals along a line with a small angular offset from the major axis. Conversely, it will overpredict the emission of the backside, which dominates the emission along the minor axis, leading to negative residuals there.

This explanation is supported by the channel maps of \co\ for Lupus 10, which clearly show the elevated emission surface (see \citealt{AGEPRO_III_Lupus}). However, it should be noted that most of the cloud absorption is also along the minor axis of this disk, which adds to the residuals seen in Figure \ref{fig: residuals}, thereby complicating their interpretation. 

A similar pattern of positive residuals along the major axis and negative ones along the minor axis can be seen in Upper Sco 8, but much less distinct than in Lupus 10. The emitting surfaces for Upper Sco 8 are also not as clearly visible in the channels (see \citealt{AGEPRO_IV_UpperSco}). Furthermore, some of the residual peaks do not align well with either axis. We note that the disk inclination reported in \citet{AGEPRO_X_substructures} from analysing the continuum emission is of $56.24^{+0.28}_{-0.29}$ deg, instead of the 16.2 deg reported here from analysing the gas-disk (Table \ref{tab: best fit Nuker profiles}). We think the flared surface of the disk might have affected our fit of the disk inclination. In general the residuals of Upper Sco 8 are much fainter than most of the other sources discussed here, making difficult to assess whether they trace real physical structures or are just due to subtracting a less than perfect fitting model. 

\emph{Upper Sco 1}: The continuum emission of Upper Sco 1 shows a large cavity, which is mirrored in several of the lines also detected for this source \citep{Sierra2024,AGEPRO_IV_UpperSco}. The \co\ emission appears to be comprised of a ring at $\sim0\farcs5$ plus a bright peak in the center. It therefore comes as no surprise that a Nuker profile is too simple a model for this particular source. For a detailed analysis of this source we refer the reader to \cite{Sierra2024}.

\begin{figure}
    \centering
    \includegraphics[width=\columnwidth]{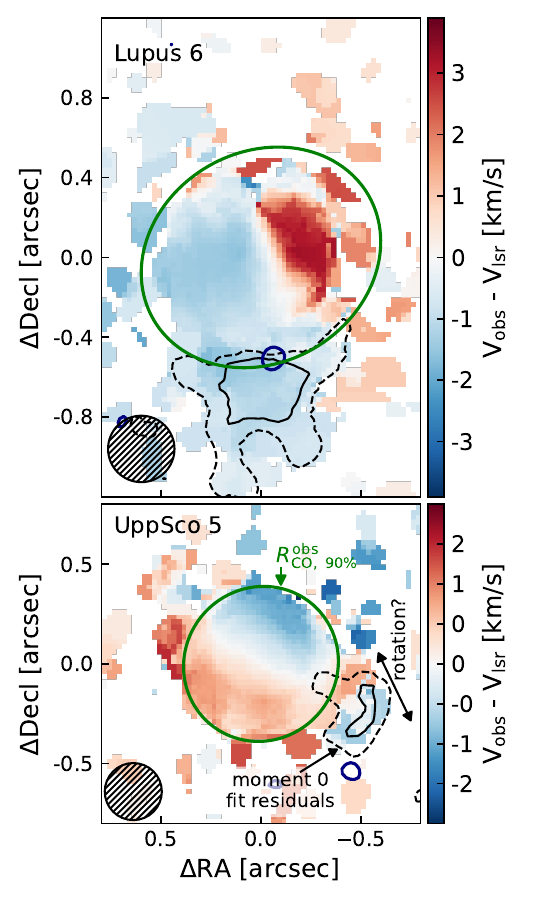}
    \caption{\label{fig: Upper Sco 5 moment 1} Moment one maps of Lupus 6 and Upper Sco 5. Black solid and dashed contours show the $5\sigma$ and $3\sigma$ contours from the moment zero fit. The green circle shows the observed \rgas\ (i.e. not from the best model). The emission inside the black contours shows tentative signs of rotation. Dark blue contours mark the location of $3\sigma$ continuum peaks in the residuals of the continuum fits from \cite{AGEPRO_X_substructures}.} 
\end{figure}

\emph{Lupus 6 and Upper Sco 5}: these two sources are the faintest and most compact sources of the seven disks examined here. Both sources show strong $(\geq 5\sigma)$ residual emission at or close to the edge of the disk. Figure \ref{fig: Upper Sco 5 moment 1} shows the location of the residual emission from the best fit model overlaid on the moment one maps of Lupus 6 and Upper Sco 5. Interestingly, the residual emission of Upper Sco 5 shows tentative signs for rotation. The residual emission of Lupus 6 at first glance seems to be only blue-shifted, but the system shows significant cloud absorption in the red part of its spectrum, so the redshifted residual emission could simply have been removed by the cloud. If the residual emission is indeed rotating that it could hint at the presence of a binary disk around an unseen companion. 
Figure \ref{fig: Upper Sco 5 moment 1} also shows the $3\sigma$ residual continuum peaks from the continuum emission analysis in \cite{AGEPRO_X_substructures}. For Lupus 6 the residual gas emission is approximately co-spatial with one of the continuum peaks.
For Upper Sco 5 relation between the continuum residual and gas asymmetry is unclear with the current sensitivity, as the potential CO rotation and continuum peak are not co-located. However, due to their close proximity, we cannot rule out that these two emission structures might be related. However, perhaps this excess gas emission is part of some large-scale cloud or infall. Deeper observations are needed to test these scenarios.
Lupus 6 is a known wide-separation binary, with a companion located at $\sim 6\farcs0$ from the primary disk (see \citealt{AGEPRO_III_Lupus} for a detailed analysis of the system) and there is no evidence in the literature that Upper Sco 5 is a binary system (see, e.g. \citealt{AGEPRO_IV_UpperSco}). Therefore while an interesting possibility, deeper, higher resolution observations are required to confirm that either system has close binary companion.

\emph{Upper Sco 7}: This disk shows some of the strongest residuals among the sources examined here. The residuals are asymmetric and dominated by the East side of the disk. Upon closer inspection this East-West asymmetry is also visible in the moment zero, one and channel maps (see \citealt{AGEPRO_IV_UpperSco}). The positive residuals at the East edge of the disk could indicate the presence of a ring, or potentially a spiral arm, on this side of the disk. Interestingly, \cite{AGEPRO_XII_mmFlare} reports the detection of a millimeter flare for this source. One of their suggested origins for this flare is a close binary, which could also explain a potential spiral in the \co\ emission. For a more detailed analysis of this source we refer the reader to \cite{AGEPRO_XII_mmFlare}.

%-------------------------------------------------------------------
\section{Conclusions}
\label{sec: conclusions}

In this work we fitted 2D Nuker and S\'ersic profiles convolved with the clean beam to \co\ J=2-1 moment zero maps of the twenty class II disks in Lupus ($\sim1-3$ Myr) and Upper Sco ($\sim2-6$ Myr) from the AGE-PRO ALMA large program. We then measure gas disk sizes (\rgasmodel) from the unconvolved best fit models, which we then compare to dust disk sizes (\rdustfrank) from \cite{AGEPRO_X_substructures}, who obtained them from fitting the continuum visibilities, to explore how the gas-to-dust size ratio evolves over time. Our conclusions are as follows: 

\begin{itemize}
    \item The best fit model gas disk sizes range from $\sim 0\farcs2$ (Lupus 9, Upper Sco 5) up to $5\farcs3$ (Lupus 10). Compared to \rgasobs, the gas disk size directly measured from the images, \rgasmodel\ is smaller by $\sim0\farcs3$, which is the typical beam major axis of the AGE-PRO \co\ 2-1 observations.

    \item The median disk size increases from $\rgasmodel = {74.5}^{+4.4}_{-7.4}$ au for the younger ($\sim1-3$ Myr) Lupus disks (range ${30}$ to ${826}$ au) to $\rgasmodel = {109.1}^{1.5}_{-3.3}$ au for the older ($\sim2-6$ Myr) Upper Sco disks (range ${30}$ to ${182}$ au). Such an increase is in line with expectations for viscously driven disk evolution but can also, at least in part, be explained by sample selection effects and survivorship biases.

    \item We find gas-to-dust size ratios (\rgasmodel/\rdustfrank) between $\sim1$ and $\sim5.5$, with a median value of $2.87^{+0.38}_{-0.36}$ consistent with previous studies. Interestingly, we find that the younger disks in Lupus have a larger median ratio $(\langle\rgasmodel/\rdustfrank\rangle) =  3.02^{+0.33}_{-0.33})$ than the older disks in Upper Sco $(\langle\rgasmodel/\rdustfrank\rangle) = 2.46^{+0.53}_{-0.38})$. This the opposite of what is expected from both dust evolution and viscous disk evolution. A possible explanation could be a combination of halted dust drift in particle traps and the effect of external photo-evaporation in Upper Sco.

    \item The best fit Nuker profiles of the larger disks ($\rgasmodel \geq 0\farcs6$) have a relatively shallow powerlaw $\gamma\approx0.3-0.9$ out to $\sim1\farcs0$ followed by a steep drop in intensity, which matches the expected shape of the intensity profile of the optically thick \co. Some of the smaller disks have steeper powerlaw slopes $(\gamma\sim1.4)$, but these disks are often only marginally resolved meaning there are limited constraints on their intensity profile.
    Three disks, Upper Sco 3, 6, and 9, have a positive inner powerlaw slope, which could indicate a gas cavity, but these cavities are not resolved in the observations.

\end{itemize}

%-------------------------------------------------------------------
\begin{acknowledgements}
L.T. and K. Z. acknowledge the support of the NSF AAG grant \#2205617. 
N.T.K. acknowledges support provided by the Alexander von Humboldt Foundation in the framework of the Sofja Kovalevskaja Award endowed by the Federal Ministry of Education and Research.
G.R. acknowledges funding from the Fondazione Cariplo, grant no. 2022-1217, and the European Research Council (ERC) under the European Union’s Horizon Europe Research \& Innovation Programme under grant agreement no. 101039651 (DiscEvol). Views and opinions expressed are however those of the author(s) only, and do not necessarily reflect those of the European Union or the European Research Council Executive Agency. Neither the European Union nor the granting authority can be held responsible for them.
P.P. and A.S. acknowledge the support from the UK Research and Innovation (UKRI) under the UK government’s Horizon Europe funding guarantee from ERC (under grant agreement No 101076489).
A.S. also acknowledges support from FONDECYT de Postdoctorado 2022 $\#$3220495.
B.T. acknowledges support from the Programme National “Physique et Chimie du Milieu Interstellaire” (PCMI) of CNRS/INSU with INC/INP and co-funded by CNES.
I.P. and D.D. acknowledge support from Collaborative NSF Astronomy \& Astrophysics Research grant (ID: 2205870).
J.M acknowledges support from FONDECYT de Postdoctorado 2024 \#3240612
C.A.G. and L.P. acknowledge support from FONDECYT de Postdoctorado 2021 \#3210520.
L.P. also acknowledges support from ANID BASAL project FB210003.
L.A.C and C.G.R. acknowledge support from the Millennium Nucleus on Young Exoplanets and their Moons (YEMS), ANID - Center Code NCN2021\_080 and 
L.A.C. also acknowledges support from the FONDECYT grant \#1241056.
All figures were generated with the \texttt{PYTHON}-based package \texttt{MATPLOTLIB} \citep{Hunter2007}. This research made use of Astropy,\footnote{http://www.astropy.org} a community-developed core Python package for Astronomy \citep{astropy:2013, astropy:2018, astropy:2022}.
\end{acknowledgements}

\software{Astropy \citep{astropy:2013, astropy:2018, astropy:2022}, 
          Numpy \citep{harris2020array},
          SciPy \citep{2020SciPy-NMeth},
          GoFish \citep{GoFish},
          bettermoments \citep{TeagueForeman-Mackey2018,Teague2019}}

%%%%%%%%%%%%%%%%%%%%%%%%%%%%%%%%%%%%%%%%%%%%%%%%%%
%  References
\bibliographystyle{aasjournal}
\bibliography{references}

%%%%%%%%%%%%%%%%%%%%%%%%%%%%%%%%%%%%%%%%%%%%%%%%%%
% Appendices
%-------------------------------------------------------------------
\begin{appendix}

\section{The considered intensity profiles}
\label{app: profile math}

A Gaussian profile\footnote{\url{https://docs.astropy.org/en/stable/api/astropy.modeling.functional_models.Gaussian2D.html}}
\begin{equation}
\label{eq: Gaussian}
    I(x,y) = A e^{ -a\left(x-x_0\right)^2 -b\left(x-x_0\right)\left(y-y_0\right) - c\left(y-y_0\right)^2}
\end{equation}

where $a,b,c$ are defined as:
\begin{align}
    a &\equiv \left(\frac{\cos^2(\theta )}{2\sigma_x^2} + \frac{\sin^2(\theta )}{2\sigma_y^2}\right)\\
    b &\equiv \left(\frac{\sin(2\theta )}{2\sigma_x^2} - \frac{\sin(2\theta )}{2\sigma_y^2}\right)\\
    c &\equiv \left(\frac{\sin^2(\theta )}{2\sigma_x^2} + \frac{\cos^2(\theta )}{2\sigma_y^2}\right)
\end{align}
Note that $\theta$ is the counter-clockwise angle measured from the x-axis.
\\

A S\'ersic profile\footnote{\url{https://docs.astropy.org/en/stable/api/astropy.modeling.functional_models.Sersic2D.html}} (e.g. \citealt{Sersic1963,GrahanDriver2005}):
\begin{equation}
\label{eq: sersic}
    I(x,y) = I_e  \exp\left( -b_n\left[\left(\frac{r(x,y)}{r_e}\right)^{\frac{1}{n}} -1\right] \right)
\end{equation}

where the radial coordinate $r$ is defined as:
\begin{align}
    r(x,y)^2 &= A^2 + \left(\frac{B}{1-ellip}\right)^2\\
    A        &\equiv (x - x_0)\cos(\theta) + (y-y_0)\sin(\theta)\\
    B        &\equiv -(x - x_0)\sin(\theta) + (y-y_0)\cos(\theta)
\end{align}
and the constant $b_n$ is defined such that $r_e$ encloses half of the total luminosity.
\\

A Nuker profile (e.g. \citealt{Lauer1995,Tripathi2017,Baes2020}):
\begin{equation}
    I(x,y)  = A \left(\frac{R}{R_b}\right)^{-\gamma} \left[1 + \left(\frac{R}{R_b}\right)^{\alpha}\right]^{\frac{\gamma-\beta}{\alpha}}.
\end{equation}
Here $R$ is the circular radius in the disk coordinate frame\footnote{e.g. \url{https://fishing.readthedocs.io/en/latest/user/fishing_coordinates.html}}, which is related to the on-sky coordinates $x_{\rm sky},y_{\rm sky}$ through
\begin{align}
R            &= \sqrt{x_{\rm disk}^2 + y_{\rm disk}^2}\\
x_{\rm proj} &= y_{\rm sky} \cos (PA) + x_{\rm sky} \sin (PA)\\
y_{\rm proj} &= \left(x_{\rm sky} \cos (PA) - y_{\rm sky} \sin (PA)\right)/\cos(i)
\end{align}
where $PA$ and $i$ are the position angle and inclination of the disk, respectively.

\section{Keplerian mask parameters}
\label{app: Keplerian mask}

This appendix contains Table \ref{tab: keplerian mask parameters}, which describes the parameters used for the Keplerian masks and other auxiliary data, like distances, disk masses, and SIMBAD source names.

%%% last updated: 10 March 2024
\begin{table*}[htb]
\centering
\caption{\label{tab: keplerian mask parameters} Keplerian mask parameters }
\def\arraystretch{1.0}%  1 is the default
\begin{tabular*}{0.92\textwidth}{ll|cccccccc}
\hline\hline
name & SIMBAD  &  $M_*$ &  dist & $v_{\rm lsr}$ & PA & inc & $z_0$ & $R_{\rm max}$ & $n_{\rm beams}$ \\
    & name &  [M$_{\rm odot}$] &  [pc] & [km/s] & [deg] & [deg] & [arcsec] & [arcsec] &  \\
\hline
Lupus 1 & Sz 65 (IK Lup) &  0.96 &  155.0 & 4.50 & 290 & 61 & 0.00 & 1.92 & 1.5\\
Lupus 2 & Sz 71 (GW Lup) &  0.77 &  155.9 & 3.70 & 36 & 48 & 0.35 & 3.30 & 1.2\\
Lupus 3 & 2MASS J16124373-3815031 &  0.80 &  160.0 & 4.60 & 198 & 51 & 0.23 & 1.44 & 2.0\\
Lupus 4 & Sz 72 (HM Lup) &  0.42 &  155.9 & 3.40 & 223 & 45 & 0.23 & 1.50 & 1.5\\
Lupus 5 & Sz 77 &  0.79 &  155.0 & 3.20 & 108 & 33 & 0.40 & 1.50 & 1.5\\
Lupus 6 & 2MASS J16085324-3914401 &  0.29 &  167.7 & 2.80 & 285 & 42 & 0.40 & 1.50 & 1.5\\
Lupus 7 & Sz 131 &  0.30 &  160.5 & 3.50 & 340 & 56 & 0.20 & 0.75 & 1.5\\
Lupus 8 & Sz 66 &  0.29 &  157.3 & 4.50 & 260 & 68 & 0.23 & 0.75 & 1.0\\
Lupus 9 & Sz 95 &  0.29 &  158.2 & 3.10 & 57 & 67 & 0.20 & 1.35 & 1.5\\
Lupus 10 & V1094 Sco &  1.10 &  153.6 & 5.42 & 110 & 50 & 0.40 & 12.00 & 2.5\\
Upper Sco 1 & 2MASS J16120668-3010270 &  0.73 &  132.0 & 4.53 & 45 & 36 & 0.20 & 3.00 & 1.5\\
Upper Sco 2 & 2MASS J16054540-2023088 &  0.20 &  138.0 & 3.00 & 49 & 56 & 0.20 & 1.50 & 1.5\\
Upper Sco 3 & 2MASS J16020757-2257467 &  0.85 &  140.7 & 4.40 & 78 & 58 & 0.20 & 1.50 & 1.5\\
Upper Sco 4 & 2MASS J16111742-1918285 &  1.20 &  137.0 & 4.24 & 256 & 76 & 0.20 & 0.60 & 1.0\\
Upper Sco 5 & 2MASS J16145026-2332397 (BV Sco) &  0.40 &  144.0 & 4.90 & 136 & 10 & 0.40 & 0.90 & 1.5\\
Upper Sco 6 & 2MASS J16163345-2521505 &  0.51 &  158.0 & 6.50 & 242 & 62 & 0.20 & 2.40 & 2.0\\
Upper Sco 7 & 2MASS J16202863-2442087 &  0.70 &  153.0 & 4.00 & 188 & 40 & 0.10 & 1.95 & 1.0\\
Upper Sco 8 & 2MASS J16221532-2511349 &  1.20 &  139.0 & 3.60 & 16 & 39 & 0.30 & 1.50 & 2.3\\
Upper Sco 9 & 2MASS J16082324-1930009 &  0.90 &  137.8 & 4.40 & 303 & 75 & 0.30 & 2.25 & 2.0\\
Upper Sco 10 & 2MASS J16090075-1908526 &  0.90 &  137.4 & 4.20 & 325 & 49 & 0.20 & 1.50 & 1.5\\
\hline\hline
\end{tabular*}
\begin{minipage}{0.85\textwidth}
\vspace{0.1cm}
{\footnotesize{\textbf{Notes:} From left to right the mask parameters are the stellar mass, distance to the source \citep{Gaia_dr3},
 the source velocity, the position angle and inclination of the disk, the aspect ratio
 at 1\farcs0 of the emission surface, the outer radius of the mask (before convolution),
 and the size of the convolution kernel to smooth the mask by (in numbers of clean beam major axes).
 For details, see \url{https://fishing.readthedocs.io/en/latest/user/api.html\#gofish.imagecube.keplerian_mask}.
Note that these parameters are only used for the Keplerian mask and should not be used as the actual stellar and disk parameters.}}
\end{minipage}
\end{table*}

\section{Comparison of Nuker and S\'ersic \rgasmodel}
\label{app: comparing Nuker and Sersic}

\begin{figure}
    \centering
    \includegraphics[width=0.9\textwidth]{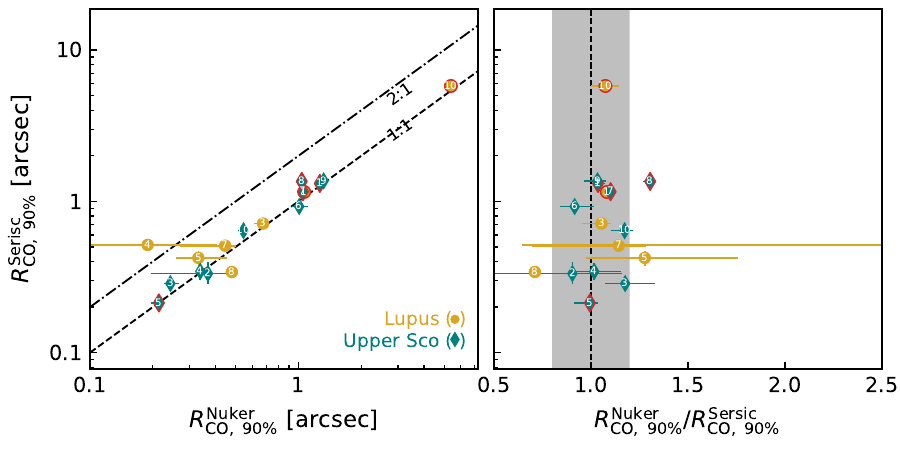}
    \caption{\label{fig: comparing Nuker and Sersic} Comparison of the \rgasmodel\ for the best fit Nuker and Sersic intensity profile. Right panel shows the ratio of the two, with the gray region showing differences up to 20\%.}
\end{figure}

In this appendix we compare the gas-disk \rgasmodel\ radii obtained with Nuker and S\'ersic profiles for the AGE-PRO sample (Fig. \ref{fig: comparing Nuker and Sersic}).

\section{$^{12}$CO 2-1 moment zero fit results}
\label{app: fit results}

\begin{figure*}
    \centering
    \begin{minipage}{\textwidth}
    \includegraphics[width=0.98\textwidth]{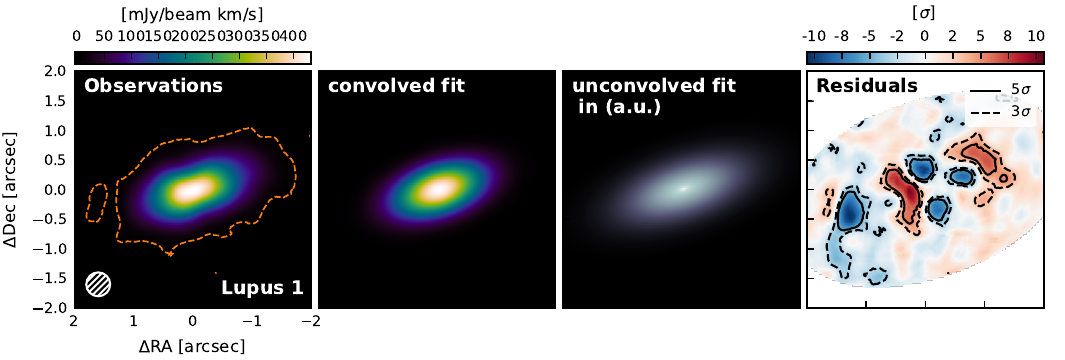}
    \end{minipage}
    \begin{minipage}{\textwidth}
    \includegraphics[width=0.98\textwidth]{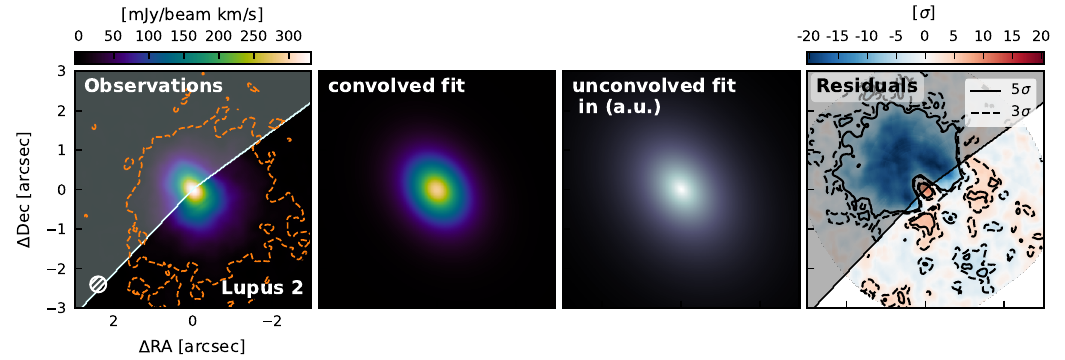}
    \end{minipage}
    \begin{minipage}{\textwidth}
    \includegraphics[width=0.98\textwidth]{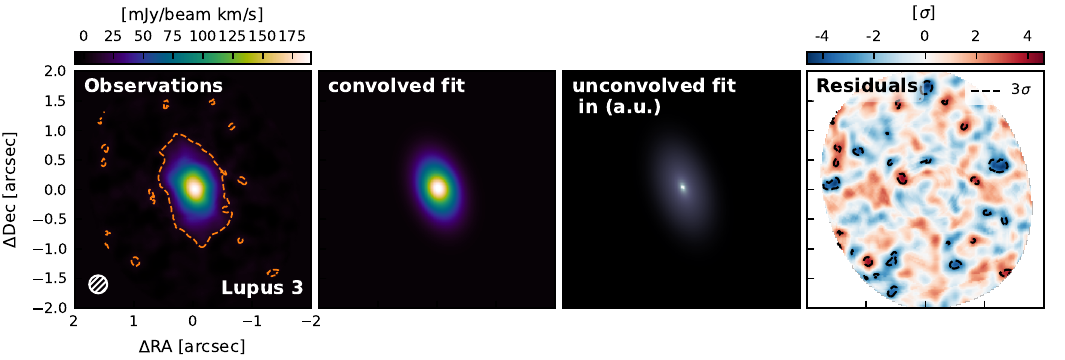}
    \end{minipage}
    \caption{\label{fig: CO moment fits Lupus 1 to 3} Results of fitting a 2D Nuker profile to the \co\ 2-1 moment zero maps of the 20 AGE-PRO sources in Lupus and Upper Sco. From left to right the panels show the observations in Jy/beam km/s, the best fit Nuker profile convolved with the clean beam Jy/beam km/s, the unconvolved best fit in arbitrary units, and the residuals obtain by subtracting convolved best fit model from the observations, expressed in units of uncertainty of the observations. The orange dashed line in the leftmost panels denotes $3\sigma$. The black dashed and solid lines in the rightmost panel shows $3\sigma$ and $5\sigma$, respectively. The black shaded region in the leftmost and rightmost panels shows the regions of the moment zero map that were excluded during the fit. Parameters of the best fit models are summarized in Table \ref{tab: best fit Nuker profiles}. }
\end{figure*}

\begin{figure*}
    \centering
    \begin{minipage}{\textwidth}
    \includegraphics[width=0.98\textwidth]{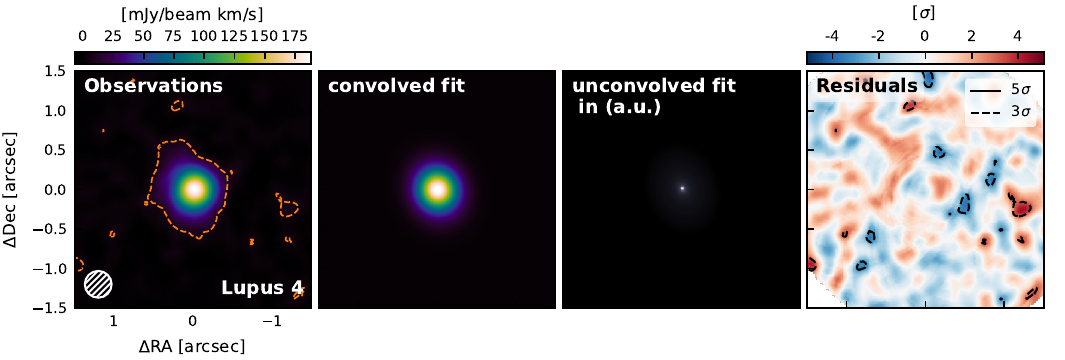}
    \end{minipage}
    \begin{minipage}{\textwidth}
    \includegraphics[width=0.98\textwidth]{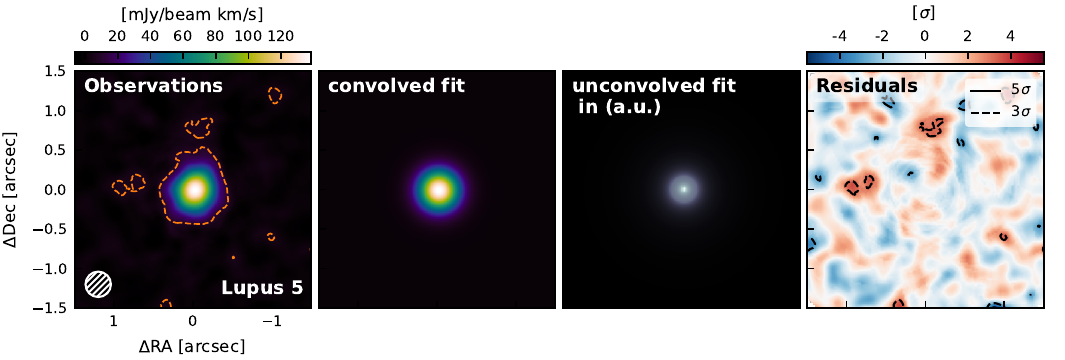}
    \end{minipage}
    \begin{minipage}{\textwidth}
    \includegraphics[width=0.98\textwidth]{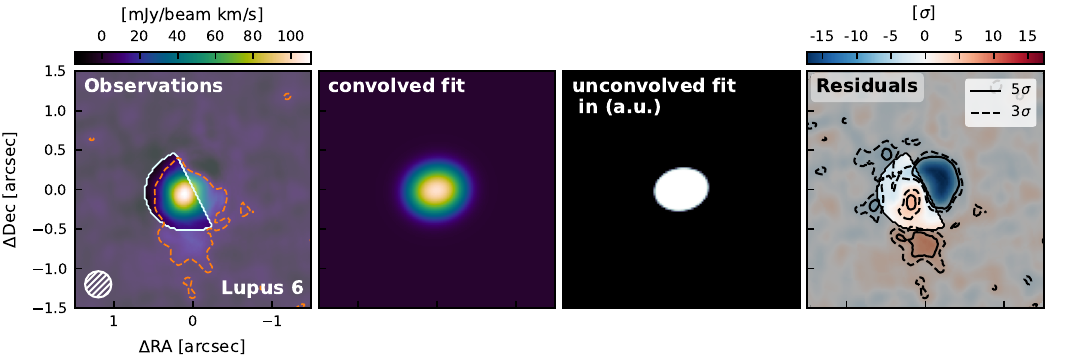}
    \end{minipage}
    \begin{minipage}{\textwidth}
    \includegraphics[width=0.98\textwidth]{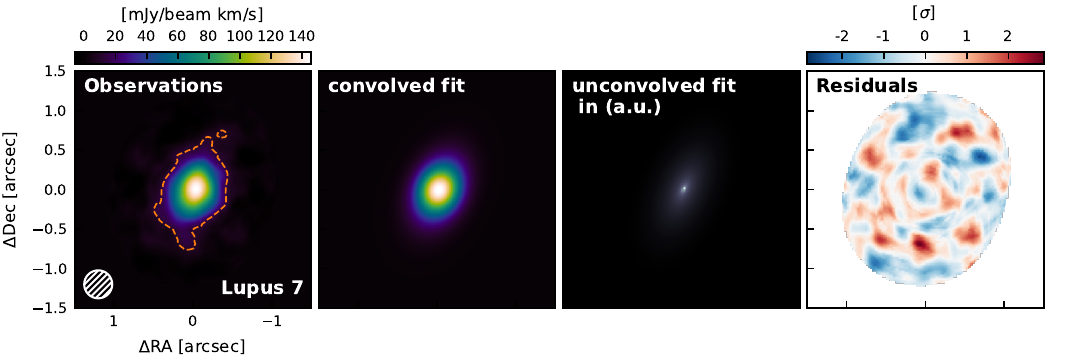}
    \end{minipage}
    \caption{\label{fig: CO moment fits Lupus 4 to 7} As Figure \ref{fig: CO moment fits Lupus 1 to 3}, but showing Lupus 4, 5, 6 and 7}
\end{figure*}

\begin{figure*}
    \centering
    \begin{minipage}{\textwidth}
    \includegraphics[width=0.98\textwidth]{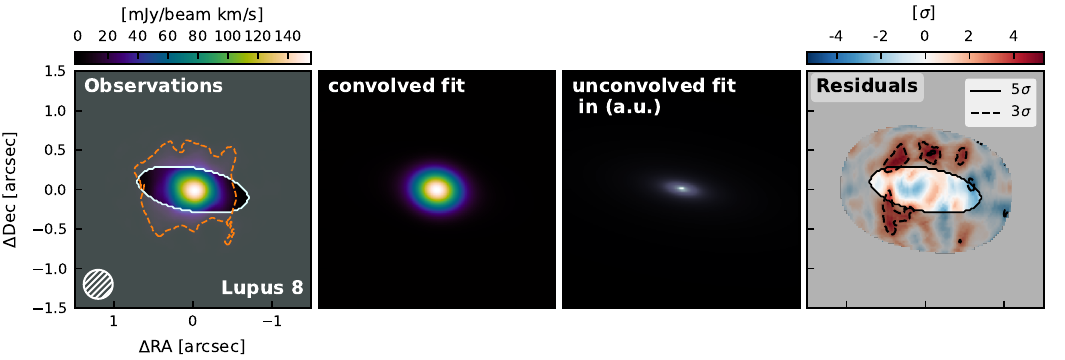}
    \end{minipage}
    \begin{minipage}{\textwidth}
    \includegraphics[width=0.98\textwidth]{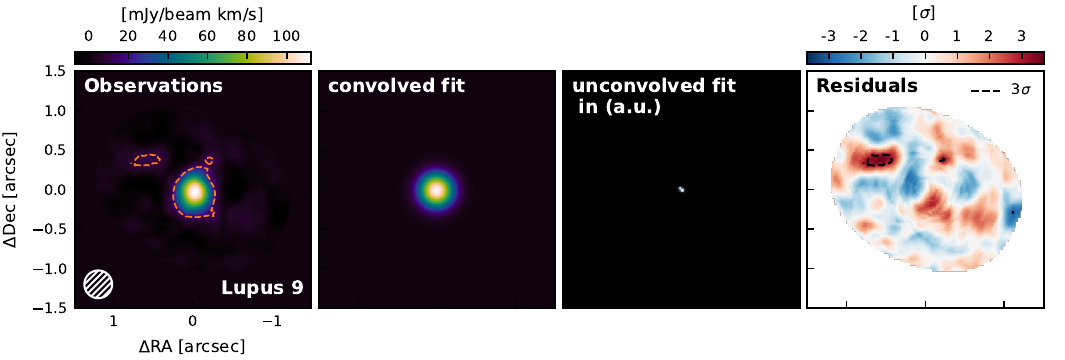}
    \end{minipage}
    \begin{minipage}{\textwidth}
    \includegraphics[width=0.98\textwidth]{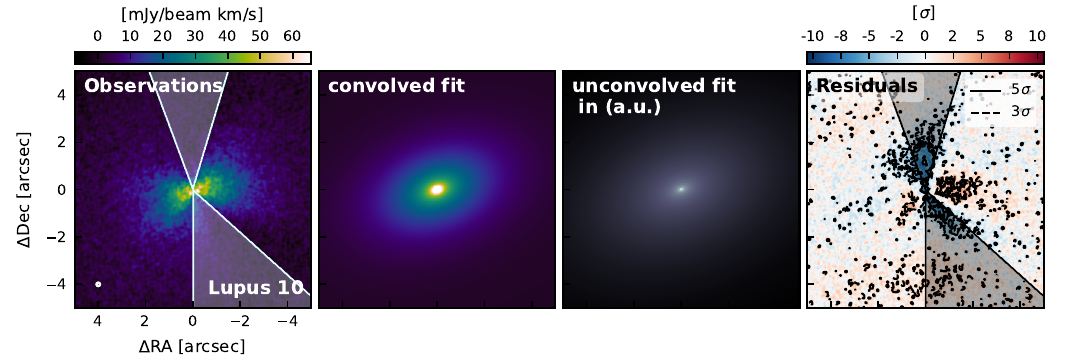}
    \end{minipage}
    \caption{\label{fig: CO moment fits Lupus 8 to 10}As Figure \ref{fig: CO moment fits Lupus 1 to 3}, but showing Lupus 8, 9, and 10}
\end{figure*}

\begin{figure*}
    \centering
    \begin{minipage}{\textwidth}
    \includegraphics[width=0.98\textwidth]{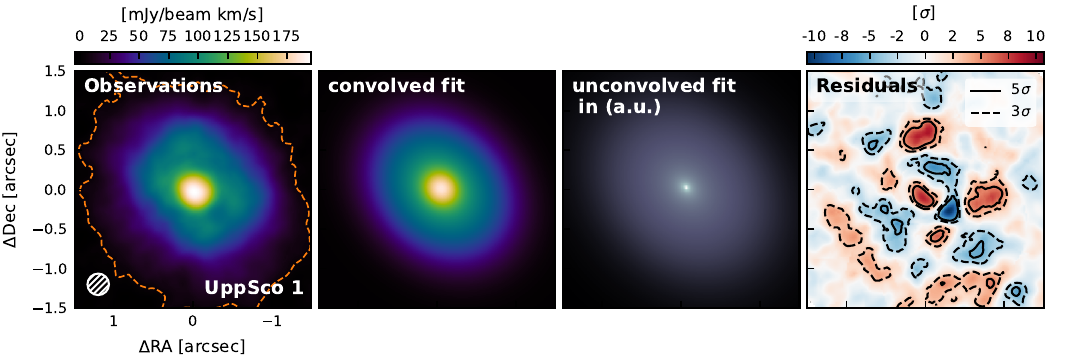}
    \end{minipage}
    \begin{minipage}{\textwidth}
    \includegraphics[width=0.98\textwidth]{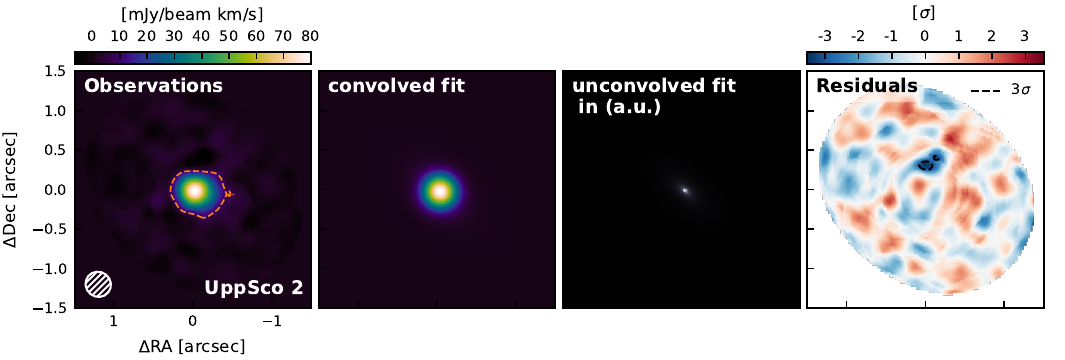}
    \end{minipage}
    \begin{minipage}{\textwidth}
    \includegraphics[width=0.98\textwidth]{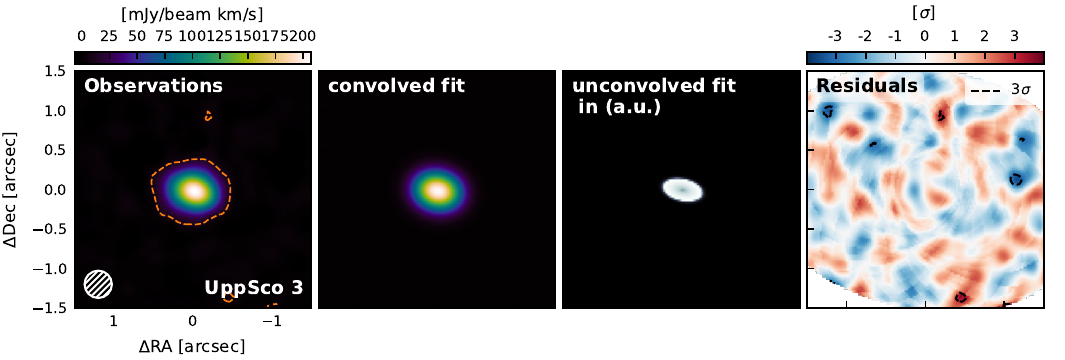}
    \end{minipage}
    \begin{minipage}{\textwidth}
    \includegraphics[width=0.98\textwidth]{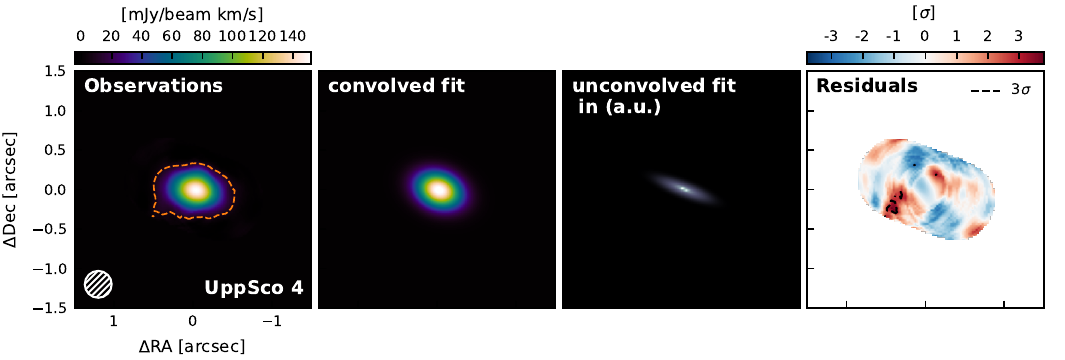}
    \end{minipage}
    \caption{\label{fig: CO moment fits UppSco 1 to 4}As Figure \ref{fig: CO moment fits Lupus 1 to 3}, but showing Upper Sco 1, 2, 3 and 4}
\end{figure*}

\begin{figure*}
    \centering
    \begin{minipage}{\textwidth}
    \includegraphics[width=0.98\textwidth]{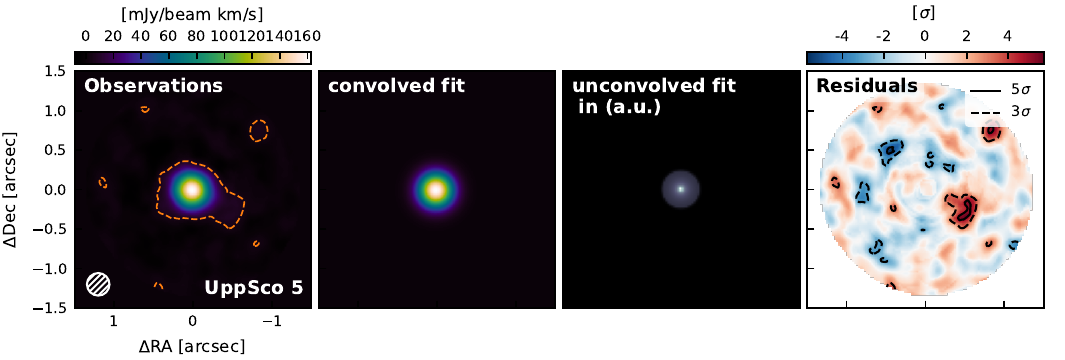}
    \end{minipage}
    \begin{minipage}{\textwidth}
    \includegraphics[width=0.98\textwidth]{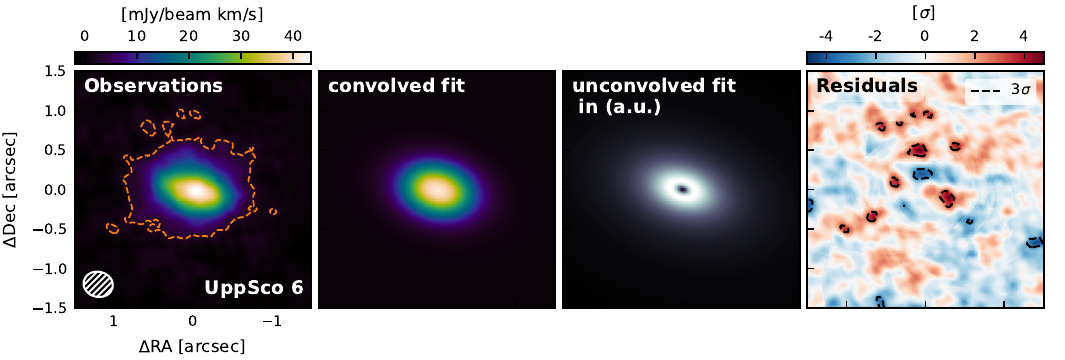}
    \end{minipage}
    \begin{minipage}{\textwidth}
    \includegraphics[width=0.98\textwidth]{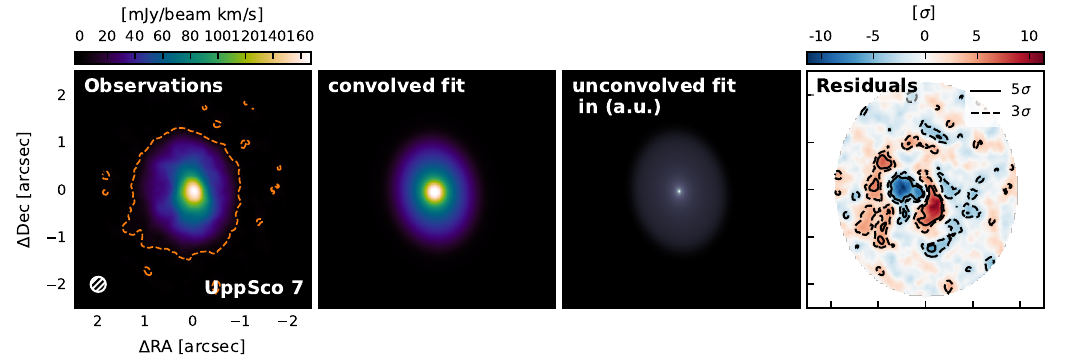}
    \end{minipage}
    \begin{minipage}{\textwidth}
    \includegraphics[width=0.98\textwidth]{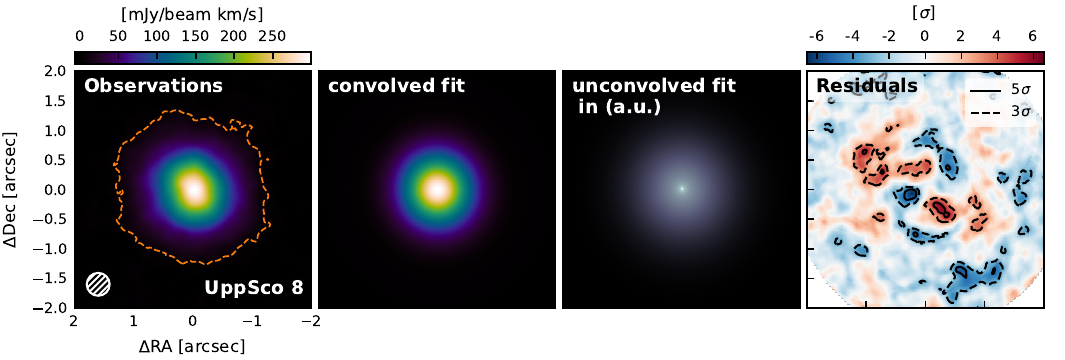}
    \end{minipage}
    \caption{\label{fig: CO moment fits UppSco 5 to 8}As Figure \ref{fig: CO moment fits Lupus 1 to 3}, but showing Upper Sco 5, 6, 7 and 8}
\end{figure*}

\begin{figure*}
    \centering
    \begin{minipage}{\textwidth}
    \includegraphics[width=0.98\textwidth]{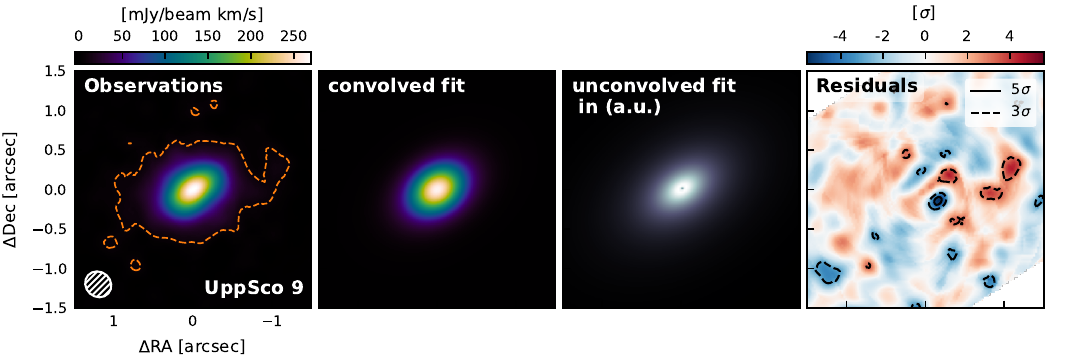}
    \end{minipage}
    \begin{minipage}{\textwidth}
    \includegraphics[width=0.98\textwidth]{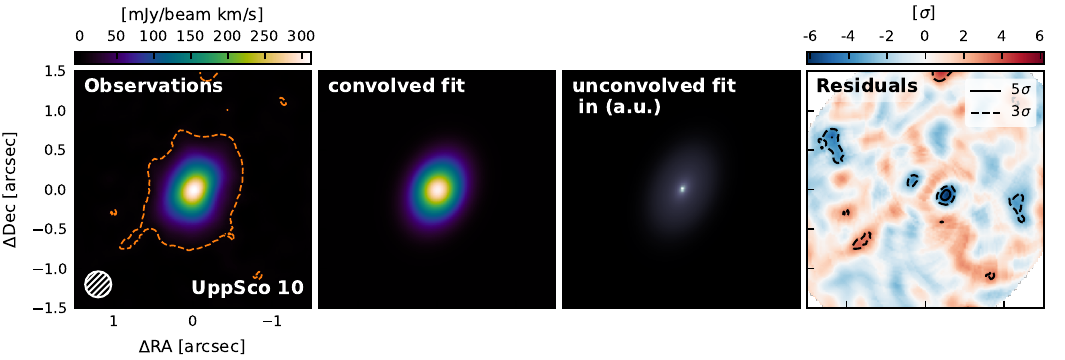}
    \end{minipage}
    \caption{\label{fig: CO moment fits UppSco 9 to 10}As Figure \ref{fig: CO moment fits Lupus 1 to 3}, but showing Upper Sco 9 and 10}
\end{figure*}

In this appendix, we show the $^{12}$CO 2-1 moment zero fit results of this work for each of the 10 Lupus (Figs. \ref{fig: CO moment fits Lupus 1 to 3}, \ref{fig: CO moment fits Lupus 4 to 7}, \ref{fig: CO moment fits Lupus 8 to 10}) and 10 Upper Sco AGE-PRO sources (Figs. \ref{fig: CO moment fits UppSco 1 to 4}, \ref{fig: CO moment fits UppSco 5 to 8}, \ref{fig: CO moment fits UppSco 9 to 10}).

\end{appendix}

\end{document}